\documentclass[twocolumn,9pt]{extarticle}
\usepackage[letterpaper,textwidth=178mm,textheight=247mm,columnsep=5mm,top=16mm]{geometry}

\newcommand{\colfigwidth}{\columnwidth}
\usepackage[T1]{fontenc}
\usepackage{textcomp}
\usepackage{lmodern}
\usepackage{graphicx}
\usepackage{xcolor}
\usepackage{comment}
\usepackage{amsmath,amssymb}
\usepackage{upgreek}
\usepackage{booktabs}
\usepackage{colortbl}   %
\usepackage{array}
\usepackage{tabularx}
\usepackage{enumitem}
\usepackage{subcaption}   %
\usepackage[skins,breakable]{tcolorbox}   %
\usepackage{xspace}
\usepackage[numbers,sort&compress]{natbib}
\usepackage{footmisc}
\usepackage[colorlinks=true,linkcolor=blue,urlcolor=blue,citecolor=blue]{hyperref}
\usepackage{authblk}
\usepackage{orcidlink}

\newcommand{\benchname}{\texttt{HEP-Ex-Bench}}
\newcommand{\bench}{\benchname\xspace}

\DeclareRobustCommand{\scpart}[1]{\scalebox{0.78}{#1}}   %
\newcommand{\hfl}{\texorpdfstring{H\scpart{IGGS}L\scpart{EP}M\scpart{ASS}}{HiggsLepMass}\xspace}
\newcommand{\hwidth}{\texorpdfstring{H\scpart{IGGS}W\scpart{IDTH}}{HiggsWidth}\xspace}
\newcommand{\lepcalib}{\texorpdfstring{L\scpart{EP}C\scpart{ALIB}}{LepCalib}\xspace}
\newcommand{\jetcalib}{\texorpdfstring{J\scpart{ET}C\scpart{ALIB}}{JetCalib}\xspace}

\newcommand{\sqrts}{\sqrt{s}}
\newcommand{\mh}{m_{\mathrm{H}}}
\newcommand{\mfl}{m_{4\ell}}
\newcommand{\mll}{m_{\ell\ell}}
\newcommand{\pt}{p_{\mathrm{T}}}
\newcommand{\Pmu}{\texorpdfstring{\ensuremath{\upmu}}{mu}\xspace}
\newcommand{\Pe}{\texorpdfstring{\ensuremath{\mathrm{e}}}{e}\xspace}
\newcommand{\Pgamma}{\texorpdfstring{\ensuremath{\upgamma}}{gamma}\xspace}

\newcommand{\Zll}{\mathrm{Z}\to\ell\ell}
\newcommand{\TeV}{\ensuremath{\,\mathrm{TeV}}\xspace}
\newcommand{\hzz}{\mathrm{H}\to\mathrm{Z}\mathrm{Z}\to 4\ell}
\newcommand{\GH}{\Gamma_{\mathrm{H}}}
\newcommand{\GSM}{\Gamma_{\mathrm{H}}^{\mathrm{SM}}}
\newcommand{\ifb}{\ensuremath{\,\mathrm{fb}^{-1}}\xspace}
\newcommand{\GeV}{\ensuremath{\,\mathrm{GeV}}\xspace}
\newcommand{\MeV}{\ensuremath{\,\mathrm{MeV}}\xspace}

\newcommand{\Sv}{S}
\newcommand{\Dtot}{\Delta_{\mathrm{total}}}
\newcommand{\Dphys}{\Delta_{\mathrm{phys}}}
\newcommand{\Dcal}{\Delta_{\mathrm{cal}}}
\newcommand{\Dgof}{\Delta_{\mathrm{gof}}}
\newcommand{\Dsyst}{\Delta_{\mathrm{syst}}}
\newcommand{\Dtruth}{\Delta_{\mathrm{truth}}}
\newcommand{\Dcomp}{\Delta_{\mathrm{compute}}}
\newcommand{\Dclosure}{\Delta_{\mathrm{closure}}}
\newcommand{\sref}{\sigma_{\mathrm{ref}}}
\newcommand{\pull}{z}
\newcommand{\sigstar}{\sigma_{\star}}
\newcommand{\zstar}{\pull_{\star}}
\newcommand{\sstar}{s_{\star}}

\definecolor{ink}{HTML}{1D1D1B}
\definecolor{muted}{HTML}{666666}
\definecolor{refplum}{HTML}{6E6E6E}
\definecolor{reffill}{HTML}{EDEDED}
\definecolor{agentamber}{HTML}{BF9000}
\definecolor{agentfill}{HTML}{FFF2CD}
\definecolor{worldblue}{HTML}{3C78D8}
\definecolor{worldfill}{HTML}{CFE3F3}
\definecolor{sealedteal}{HTML}{3A781D}
\definecolor{sealedfill}{HTML}{D9EAD3}
\definecolor{evalcrimson}{HTML}{990001}
\definecolor{evalfill}{HTML}{F5CCCC}
\definecolor{slategrey}{HTML}{6E6E6E}
\definecolor{slatefill}{HTML}{EFEFEF}
\definecolor{stagebg}{HTML}{FAFAFA}

\long\def\todo#1{{\bfseries\color[rgb]{0.70,0.12,0.62}[TODO: #1]}}%

\newcommand{\grouprule}{\arrayrulecolor{black!35}\specialrule{0.3pt}{\aboverulesep}{\belowrulesep}\arrayrulecolor{black}}

\newcommand{\promptfont}{\renewcommand{\ttdefault}{zi4}\footnotesize\ttfamily\raggedright\setlength{\parskip}{4pt}}

\makeatletter
\newcounter{email}
\gdef\email@list{}
\DeclareRobustCommand{\email@add}[1]{\stepcounter{email}\alph{email}%
  \protected@xdef\email@list{\email@list\noexpand\email@box{\the\c@email}{\alph{email}}{#1}}}
\newcommand{\email}[1]{\texorpdfstring{\email@add{#1}}{}}
\newcommand{\lead}{*}
\newcount\email@half
\DeclareRobustCommand{\email@box}[3]{\ifnum#1>\email@lo\relax\ifnum#1>\email@hi\relax\else\hspace{1em}\textsuperscript{#2}\href{mailto:#3}{#3}\par\fi\fi}
\newcommand{\email@col}[2]{\def\email@lo{#1}\def\email@hi{#2}\parbox[t]{0.5\columnwidth}{\raggedright\email@list}}
\newcommand{\emailnote}{\begingroup\long\def\@makefntext##1{\noindent ##1}\def\@thefnmark{}\def\Hy@footnote@currentHref{emailnote}%
  \email@half=\numexpr\value{email}/2\relax %
  \@footnotetext{\hspace{1em}\textsuperscript{*}Lead authors\par\noindent
    \email@col{0}{\the\email@half}\email@col{\the\email@half}{\value{email}}}\endgroup}
\makeatother
\definecolor{dgreen}{rgb}{0,0.5,0.1}\DeclareRobustCommand{\deltodo}[1]{[removed TODO: #1]}\DeclareRobustCommand{\ddel}[1]{{\color{red}\let\todo\deltodo #1}}

\title{\benchname: Evaluating Agentic Performance in \\ High Energy Physics Experiment Data Analyses with Sealed Worlds}

\author[1,2,\email{novaka@mit.edu},\lead]{Andrzej Nov\'ak\,\orcidlink{0000-0002-0389-5896}}
\author[1,2,\email{yiyang26@mit.edu},\lead]{Yiyang Zhao\,\orcidlink{0009-0000-2290-1828}}
\author[1,2,\email{dolores.garcia@cern.ch}]{Dolores Garcia\,\orcidlink{0000-0002-0120-8757}}
\author[1,2,\email{emoreno@mit.edu}]{Eric A. Moreno\,\orcidlink{0000-0001-5666-3637}}
\author[3,\email{jkrupa@slac.stanford.edu}]{Jeffrey Krupa\,\orcidlink{0000-0003-0785-7552}}
\author[4,\email{vbresopla@fas.harvard.edu}]{V\'ictor Bres\'o-Pla\,\orcidlink{0000-0002-0240-2738}}
\author[1,2,\email{sambt@mit.edu}]{Samuel Bright-Thonney\,\orcidlink{0000-0003-1889-7824}}
\author[5,\email{l.heinrich@tum.de}]{Lukas Heinrich\,\orcidlink{0000-0002-4048-7584}}
\author[3,\email{makagan@slac.stanford.edu}]{Michael Kagan\,\orcidlink{0000-0002-3386-6869}}
\author[1,2,\email{pcharris@mit.edu}]{Philip Harris\,\orcidlink{0000-0001-8189-3741}}
\affil[1]{Department of Physics, Massachusetts Institute of Technology, Cambridge, MA, USA}
\affil[2]{NSF AI Institute for Artificial Intelligence and Fundamental Interactions (IAIFI)}
\affil[3]{SLAC National Accelerator Laboratory, Menlo Park, CA, USA}
\affil[4]{Department of Physics, Harvard University, Cambridge, MA, USA}
\affil[5]{Department of Physics, Technical University of Munich, Garching, Germany}
\date{}

\usepackage{lineno}
\usepackage{stfloats} %
\begin{document}
    \maketitle
\emailnote

\begin{abstract}
Agentic systems are increasingly being explored for automating experimental high-energy physics measurements, but the correctness of their results can today be reliably evaluated only via extensive expert review. To address this, we propose a method to perform detailed quantitative evaluations of agent-driven analyses through the use of blind analysis and injection tests, which are critical in a typical expert review. Agents develop a data analysis pipeline on a public dataset, while the evaluation is performed on programmatically generated sealed private datasets where signals are injected and simulation settings varied within realistic detector uncertainties. By requiring agents to provide a single workflow that is evaluated across these ``sealed worlds'', we are able to test the accuracy and robustness of the workflows. We demonstrate the approach on four tasks built from CMS open data and present it as a way to reliably benchmark scientific work in experimental high-energy physics. At the time of writing, the best-performing model, Gemini 4 Argon, reaches 60\% of the score of the expert reference analyses on average, ahead of GPT 6 Astra at 38\% and Claude Opus 5.5 at 31\%. In individual sessions on one of the tasks, Gemini 4 Argon exceeds the reference. The benchmark is hosted at \url{https://hep-ex-bench.github.io}.
\end{abstract}

\vspace{0.3em}
\section{Introduction}
\begin{figure}[!t]\centering
\includegraphics[width=\columnwidth]{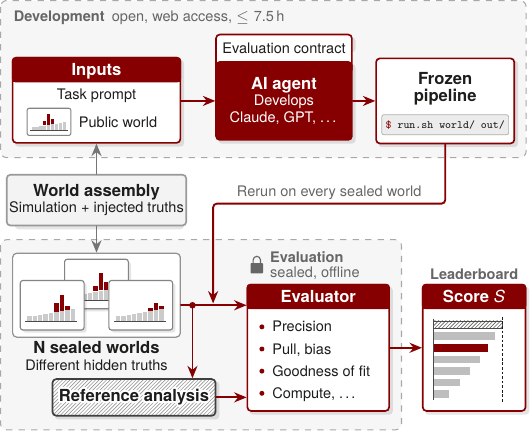}
\caption{The sealed-world benchmarking scheme. The agent works from the task prompt and a public world, and delivers a pipeline. The pipeline is then frozen and rerun on sealed worlds with different hidden truths. The evaluator scores the results against the truths and against an expert reference analysis of the same worlds. The worlds are assembled beforehand from simulation with injected truths (Fig.~\ref{fig:flow}).}
\label{fig:schematic}
\end{figure}

Scientific research is moving toward systems that execute parts of the process autonomously through artificial intelligence (AI) agents built on large language models (LLMs). In high-energy physics (HEP), agents can now carry out substantial parts of the elaborate data analysis pipeline, from event selection and background modeling to the construction of a full likelihood~\cite{jfc2026,Gendreau-Distler:2025fsj,he2026drsai}. The question is therefore no longer whether agents can analyze data, but how the quality and reliability of their analyses can be evaluated.

An analysis that runs to completion is not necessarily correct. The true result on observed data is unknown, and an agent that reproduces a published measurement may simply have recalled or looked up the accepted value~\cite{xu2024contamination,liang2025swebenchillusion}. Expert review does not scale, and a language-model evaluation~\cite{discoverphysics2026} can be exploited into accepting pipelines that would fail such a review~\cite{zheng2025cheating}.

Benchmarks are therefore needed that have a reproducible scoring methodology and are characteristic of real physics analyses, with truths isolated from the web and from training data. These benchmarks should evaluate both the precision of a measurement and whether the analysis methods are correct. To address this gap we propose sealed-world benchmarking, which follows two established methods of experimental physics, blind analysis~\cite{klein_roodman_2005} and injection tests~\cite{biwer_ligo_injections_2017}. In a blinded mode, the analysis is tested on pseudodata with a known injected truth before the signal region of the observed data is analyzed. A reliable analysis is one that recovers the injected truth within its quoted uncertainty~\cite{cowan_asymptotic_2011}.

In \bench, the High-Energy Physics Experiment Benchmark, we present a mechanism for reliable evaluation of AI agents through sealed-world injections. An agent develops an analysis once on a public dataset (public world), built from simulated samples in the format of real analysis input; the agent then delivers a pipeline that aims to produce unbiased physics results over a wide range of physics outcomes and detector behaviors. The pipeline is frozen and executed in isolation on many varied sealed worlds. These sealed worlds are built with the full detector simulation of an experiment, here the open simulated data of the CMS experiment~\cite{cms_detector_2008}. The sealed worlds differ in their parameters of interest\footnote{The quantities that an analysis sets out to measure, such as the Higgs boson mass.} and in their detector response, which are hidden from the agent. The evaluator harness scores the results against the hidden truths and against an expert reference analysis of the same worlds, and checks the method programmatically, for example the goodness of fit of the delivered model. Figure~\ref{fig:schematic} shows the scheme. A memorized value, a fit tuned to the public world, or an understated uncertainty scores poorly on the sealed worlds. A Higgs boson mass of 125\GeV, for example, is wrong in a world where it is 115\GeV. In contrast, a sound analysis recovers the truths with uncertainties consistent with the sealed-world variations.

\label{sec:intro}

 Section~\ref{sec:related} reviews related work, and Sec.~\ref{sec:evaluation} defines the sealed-world evaluation, the failure modes it is designed to catch, and the score. Section~\ref{sec:benchmarks} presents four example tasks, and Sec.~\ref{sec:datasets} describes how their worlds are built from CMS simulation~\cite{cms_od_dy,cms_od_zz,cms_od_ggzz_4e,cms_od_ggzz_4mu,cms_od_ggzz_2e2mu,cms_od_tt,cms_od_premix}. Section~\ref{sec:results} validates the expert reference analyses and the evaluator and reports a campaign of agents, Sec.~\ref{sec:discussion} discusses the results and the limits of the method, and Sec.~\ref{sec:conclusion} discusses implications and way forward for the field.

\section{Related work}
\label{sec:related}

Recent work on AI agents has largely focused on the construction of agentic workflows. \bench addresses their evaluation. Evaluation can be performed against a published value, a hidden reference or an injected truth.
The papers summarized below are grouped in that order, from the weakest to the strongest ground truth, and are compared in Table~\ref{tab:positioning}.

The most direct evaluation of an agentic workflow checks its result against published results. PaperBench asks agents to replicate machine-learning papers, runs their code again and compares the agents with human experts~\cite{starace2025paperbench}, and PRBench scores how well agents reproduce the results of published physics papers~\cite{qiu2026prbench}. In HEP, Gendreau-Distler et al.\ compare many language models on a Higgs-boson analysis of ATLAS Open Data~\cite{Gendreau-Distler:2025fsj}, Dr.Sai measures branching fractions on the full simulation of the BESIII experiment~\cite{he2026drsai}, and Collider-Bench, the work closest to this paper in its tasks, asks agents to rebuild published LHC~\cite{lhc_machine_2008} searches and scores the signal yields that they predict~\cite{colliderbench}. A published value, however, cannot reliably tell a robust analysis apart from a recalled one. Only Dr.Sai reports uncertainties, and it checks them through pulls (deviations from the truth in units of the reported uncertainty) on a single simulated sample instead of scoring them.

 A stronger evaluation compares the result with a reference that the agent cannot see. MLE-bench scores machine-learning engineering against test sets that the agents do not see but that are derived from public data~\cite{chan2024mlebench}, HEPTABench evaluates agents repeatedly on HEP tasks with hidden reference values~\cite{menzo2026heptabench}, and DiscoverPhysics asks agents to discover the laws of motion of simulated worlds, tests them on held-out trajectories and grades their explanations with a language model~\cite{discoverphysics2026}. A hidden reference guards against recall, but each task has one fixed reference and no uncertainty is scored, so an unsound analysis cannot be told from a fluctuation.

Lastly, evaluation can be performed against an injected truth, grading both the measured values and their uncertainties. The HEP data challenges, including HiggsML, the LHC Olympics and Dark Machines, and the HDR Gravitational Wave Anomaly Challenge, all construct evaluation methods for finding hidden signals in simulated collision events or gravitational wave data~\cite{higgsml2014,Kasieczka:2021xcg,darkmachines,campolongo2025buildingmachinelearningchallenges}. FAIR Universe, the closest to this work in its scoring, runs every submitted model on pseudo-experiments with re-drawn signal strengths and systematic shifts, then scores the coverage and magnitude of the reported uncertainties~\cite{fairuniverse2024}. These challenges, however, evaluate specific methods rather than agents that carry out complete analyses, and their data are precomputed features of a single analysis, so the task is statistical inference alone.

\bench combines the elements of these three groups (Table~\ref{tab:positioning}). As in the first group, the agent develops a whole analysis. As in FAIR Universe, the truths are hidden and re-drawn for every evaluation and the reported uncertainties are scored. In addition, the statistical model that the agent delivers is checked, for example for its goodness of fit, so an analysis is scored on its method as well as on its result.

\begin{table}[!t]
\centering\scriptsize
\setlength{\tabcolsep}{0pt}
\newcommand{\rh}[1]{\rotatebox{90}{\parbox{14.5mm}{\raggedright\baselineskip=6.5pt #1}}}
\newcolumntype{Q}{>{\centering\arraybackslash}p{10.2mm}}\newcolumntype{V}{>{\centering\arraybackslash}p{11.7mm}}\newcommand{\no}{\textcolor{black!45}{$\times$}}\newcolumntype{N}{>{\centering\arraybackslash}p{6.3mm}}
\begin{tabular*}{\columnwidth}{@{}l@{\extracolsep{\fill}}QVNQNNQ@{}}
\toprule
 & \rh{Truth} & \rh{Re-run on\\new data} & \rh{Uncertainty\\scored} & \rh{Full detector\\simulation} & \rh{Expert\\reference} & \rh{Complete\\analysis} & \rh{Other\\analyses} \\
\midrule
PaperBench~\cite{starace2025paperbench} & \no & \mbox{same data} & \no & N/A & \checkmark & N/A & N/A \\
Gendreau-Distler~\cite{Gendreau-Distler:2025fsj} & \no & \no & \no & real data & \no & \no & \no \\
Dr.Sai~\cite{he2026drsai} & \no & \no & \no & \checkmark & \no & \checkmark & \no \\
Collider-Bench~\cite{colliderbench} & \no & \no & \no & fast sim. & \checkmark & \no & searches \\
\grouprule
MLE-bench~\cite{chan2024mlebench} & \no & \no & \no & N/A & \checkmark & N/A & N/A \\
HEPTABench~\cite{menzo2026heptabench} & \no & \no & \no & \no & \no & \no & \checkmark \\
DiscoverPhysics~\cite{discoverphysics2026} & hidden & \checkmark & \no & N/A & \no & N/A & N/A \\
\grouprule
FAIR Universe~\cite{fairuniverse2024} & re-drawn & \checkmark & \checkmark & fast sim. & \no & \no & \no \\
\grouprule
\bench{} & re-drawn & \checkmark & \checkmark & \checkmark & \checkmark & \checkmark & \checkmark \\
\bottomrule
\end{tabular*}
\caption{Comparison with related benchmarks, agents and data challenges. A hidden truth is unknown to the agent and not on the web, and a re-drawn one changes with every test. Re-run on new data means that the submitted code is executed on data it was not developed on. Uncertainties are scored when both too small and too large ones are penalized. An expert reference is scored like the agents. A full detector simulation tracks every particle through the detector, while a fast one parametrizes its response and is therefore inherently less representative. A complete analysis runs from the reconstructed events to the final statistical result, and the last column marks schemes that extend to other kinds of analysis. }
\label{tab:positioning}
\end{table}

\section{Sealed-world evaluation}
\label{sec:evaluation}

Since the truth in experimentally collected data is unknown, an analysis in HEP is conventionally developed and calibrated on simulation, and tested with injected signals~\cite{cowan_asymptotic_2011}. The application to the collected data is only performed once the method is fixed~\cite{klein_roodman_2005}. \bench extends this practice with sealed worlds, simulated datasets that emulate experimentally collected data and whose true values are chosen by the evaluator and hidden from the tested agent (their construction is described in Sec.~\ref{sec:modifications}). An analysis is tested on many sealed worlds, each with a different injected truth, so that only a method that holds across these variations recovers the injected truths. The scheme applies to any measurement and is illustrated by four tasks (Sec.~\ref{sec:benchmarks}).

An agent works on a task in one development session, in this paper a single run of at most 7.5\,h in a sandbox. In it, the agent is given
\begin{itemize}
\item a prompt in which the task is stated,
\item the evaluation contract, in which the deliverables, output formats, and a self-check are fixed while as little task-specific information as possible is revealed (App.~\ref{app:contract}),
\item the public world -- pseudodata in the format of real analysis inputs, and
\item the nominal simulation (with default detector settings).
\end{itemize}
The agent develops a pipeline with web access to look up published methods. Once delivered, the pipeline is frozen and rerun offline, in isolation, on the sealed worlds (Fig.~\ref{fig:schematic}). A submission that breaks the contract is not evaluated.

\subsection{Evaluation on many sealed worlds}
\label{sec:multiworld}

\begin{figure}[!t]\centering
\includegraphics[width=\colfigwidth]{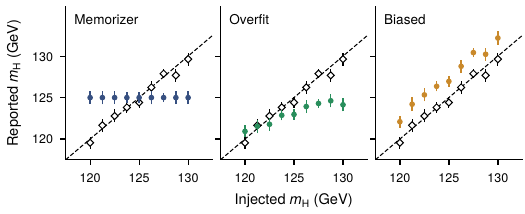}
\caption{Toy analyses that many sealed worlds expose: the reported Higgs boson mass $\mh$ against the injected one on nine worlds, with the expert reference analysis as open diamonds.}
\label{fig:toy_threats}
\end{figure}

In every task, an analysis is scored by its results on many sealed worlds. Three ways in which an analysis can look right without being right are guarded against:
\begin{itemize}
\item Knowing the answer. A value is recalled from the literature or from training data, or read from the truths, such as a 125\GeV~Higgs boson mass reported on every world where the mass is varied (Fig.~\ref{fig:toy_threats}, left). The sealed truths are drawn anew and are on no web page, and the isolated runs and the session transcripts are checked for access to them.
\item Overfitting to the public world. An analysis is tuned on the one public dataset provided to the agent, and is right for that world but biased toward its value on the others (Fig.~\ref{fig:toy_threats}, center). The pipeline is frozen before it runs on any sealed world, and those worlds differ in their parameters of interest, background simulation, and detector response; only an analysis with full detector, background and signal variations will result in an unbiased result.
\item Guessing. On a single world, a guess can agree with the truth, and a sound analysis can miss it through a fluctuation. The two are separated on many worlds, as shown by the toy study in App.~\ref{app:toy}.
\end{itemize}
 Misstated uncertainties, a bias shared by every world (Fig.~\ref{fig:toy_threats}, right) and unsound methods are caught by the score itself, through the scoring rule, the combination of the worlds and the quality terms (Sec.~\ref{sec:scoring}).
 The checks are verified with deliberately flawed and cheating submissions, whose scores are given in App.~\ref{app:adversarial}.

\subsection{Score}
\label{sec:scoring}\label{sec:reference}

Every analysis is compared with an expert reference based on published methods that recover the injected truths on sealed worlds. The evaluation score measures the accuracy of the measurement (Sec.~\ref{sec:measurement}) and quality terms such as the goodness of fit of the statistical model, computational efficiency, and the treatment of systematic uncertainties (Sec.~\ref{sec:quality}). 

\subsubsection{Measurement}
\label{sec:measurement}

The measurement score compares the analysis results against the reference performance on $N$ sealed worlds. On each world $i$ the analysis produces a value $x_i$ and an uncertainty $\sigma_i$ for a measured quantity $p$. The score for each world is then computed as the difference between the negative log-likelihood of the injected truth $t_i$ under a Gaussian of mean $x_i$ and width $\sigma_i$ between the analysis and the reference 
\begin{equation}
  \Delta_i = \underbrace{\ln\frac{\sigma_i}{\sigma_{\mathrm{ref},i}}}_{\text{precision}} + \underbrace{\tfrac12\,(\pull_i^2 - 1)\vphantom{\frac{\sigma_i}{\sigma_{\mathrm{ref},i}}}}_{\text{pull}} ,\qquad \pull_i=\frac{x_i-t_i}{\sigma_i} ,
  \label{eq:pieces}
\end{equation}
where $\sigma_{\mathrm{ref},i}$ is the reference uncertainty and $\pull_{\mathrm{ref},i}=1$. A $\Delta_i=0$ then implies that the analysis is as good as the reference. This is the logarithmic score, a \emph{strictly proper} scoring rule~\cite{good_1952,gneiting_raftery_2007} (cf.\ Ref.~\cite{fairuniverse2024}, Sec.~\ref{sec:related}). The precision piece is negative for an analysis more precise than the reference, and the pull piece averages to zero for accurate uncertainties, so the expected penalty is lowest at unit pull width, $\langle\pull_i^2\rangle=1$. There is no floor and no cap: at reference precision a pull of 1 gives no penalty, a pull of 2 a penalty of 1.5, and a pull of 3 a penalty of 4. For a quantity whose likelihood is not Gaussian, such as the signal strength in \hfl, the same rule is applied to the delivered likelihood itself, and Eq.~\eqref{eq:pieces} is its special case (App.~\ref{app:likelihood}).

 A world has no measurement when the pipeline fails or its output does not determine the quantity. It is penalized as if the nominal value stated in the task had been reported, with the spread of the values the task allows as its uncertainty. This is the answer of an analysis that learned nothing from the data. For the mass in \hfl this penalty is about 4.5, that of a pull of 3 (App.~\ref{app:missing}).

 On a single world, a bias that is small compared with the uncertainty is penalized in the same way as a fluctuation of the same size. In order to account for shared biases in the score, the residuals $x_i-t_i$ are combined by inverse variance and their combination is added to the $\Delta_i$ sum,
\begin{equation}
  \Delta_p = \underbrace{\frac1N\sum_{i=1}^{N}\Delta_i}_{\text{each world}} + \underbrace{\frac{\zstar^2-1}{2N}}_{\text{shared bias}} ,
  \label{eq:bias}
\end{equation}
where $\zstar=\sigstar\sum_i(x_i-t_i)\,\sigma_i^{-2}$ is the pull of the combination and $\sigstar=(\sum_i\sigma_i^{-2})^{-1/2}$ its uncertainty, summed over the worlds with a measurement. A bias shared by all worlds grows in $\zstar$ like $\sqrt{N}$, while an accurate analysis has $\langle\zstar^2\rangle=1$ and no penalty on average. For a task with one quantity, $\Delta_p$ is its measurement penalty $\Dphys$.

\subsubsection{Quality}
\label{sec:quality}

 Problems of the method that are not visible in the measurement directly are penalized by quality terms, modeled on the checks a reviewer would make. Each is a penalty $\Delta_k$ that is zero when the check passes, and enters the total penalty at a weight $w_k<1$ (Table~\ref{tab:quality}, App.~\ref{app:quality}),
\begin{equation}
  \Dtot = \underbrace{\Dphys\vphantom{\sum_k}}_{\text{measurement}} + \underbrace{\sum_k w_k\,\Delta_k}_{\text{quality}} .
  \label{eq:dtot}
\end{equation}
\begin{table}[!t]\centering\footnotesize
\begin{tabularx}{\columnwidth}{@{}l@{\hspace{0.6em}}c@{\hspace{0.6em}}X@{\hspace{0.6em}}l@{}}
\toprule
Term & $w_k$ & What is checked & Tasks \\
\midrule
Goodness of fit, $\Dgof$ & 0.30 & model against the pseudodata & fit \\
Detector calibration, $\Dcal$ & 0.10 & logged detector effects & fit \\
Truthfulness, $\Dtruth$ & 0.50 & report against a model refit & fit \\
Kinematic closure, $\Dclosure$ & 0.50 & $\mathrm{Z}$ peak after the corrections & cal. \\
Completeness, $\Dsyst$ & 0.10 & standard systematics modeled & both \\
Compute, $\Dcomp$ & 0.10 & wall time against the reference & both \\
\bottomrule
\end{tabularx}
\caption{Quality terms of Eq.~\eqref{eq:dtot} and their weights $w_k$.}
\label{tab:quality}
\end{table}

 The weights follow typical review priorities in a collaboration. Each is below one, so that no check outweighs the measurement. The detector-calibration term scores every detector effect the analysis records like a measurement, against the injected one, and the goodness-of-fit term penalizes a delivered model that does not describe the pseudodata well. A false discovery is penalized by the measurement term, since a world may contain no signal.

\subsubsection{Final score}
\label{sec:composition}

The final score $\Sv$ of a task and its ratio to the score of the reference on the same worlds are
\begin{equation}
  \Sv = \mathrm{e}^{-\gamma\,\Dtot} ,\qquad R=\frac{\Sv}{\Sv_{\mathrm{ref}}} = \prod_k \mathrm{e}^{-\gamma\,w_k(\Delta_k-\Delta_k^{\mathrm{ref}})} ,
  \label{eq:score}
\end{equation}
with $\gamma=2$, the product over the terms of Eq.~\eqref{eq:dtot} and $w_k=1$ for the measurement term. Results are reported as $R$ in percent. The choice of $\gamma$ puts the score in units of data: with accurate uncertainties and no quality penalty, $\Sv\approx\big(\frac{\sref}{\sigma}\big)^2=f$ is the fraction of the data on which the reference reaches the same precision, so $\Sv=50\%$ reads as an analysis as good as the reference on half the data.

 With several measured quantities, such as the mass and the signal strength in \hfl, each has its own penalty $\Delta_p$ and the scores $\mathrm{e}^{-\gamma\,\Delta_p}$ are averaged (App.~\ref{app:quantities}). A missing or far-off quantity loses at most its own share and leaves the others untouched.

The score is nominally a random variable, since it depends on the draw of the sealed worlds and on the statistical fluctuations of their pseudodata (App.~\ref{app:bands}).

\section{Tasks}
\label{sec:benchmarks}

The four tasks are based on analyses carried out in practice at the LHC, two measurements (fits) and two calibrations (Table~\ref{tab:tasks}). Each requires the full chain of a real analysis, from the calibration of the physics objects to a statistical model with its systematic uncertainties, and each has an expert reference analysis that follows a published CMS method (App.~\ref{app:closure}).

\begin{table}[!t]\centering\footnotesize
\begin{tabularx}{\columnwidth}{@{}l@{\hspace{0.6em}}l@{\hspace{0.6em}}>{\raggedright\arraybackslash}X@{\hspace{0.6em}}r@{}}
\toprule
Task & Type & Result & App. \\
\midrule
\hfl & fit & mass $\mh$ and signal strength $\mu$ & \ref{app:ref_hfl} \\
\addlinespace[2pt]
\hwidth & fit & width ratio $\xi=\GH/\GSM$ and on-shell signal strength $\mu_{\mathrm{on}}$ & \ref{app:ref_hwidth} \\
\addlinespace[2pt]
\lepcalib & cal. & \Pe and \Pmu scale, resolution and efficiency corrections & \ref{app:ref_lepcalib} \\
\addlinespace[2pt]
\jetcalib & cal. & jet energy scale and resolution corrections & \ref{app:ref_jetcalib} \\
\bottomrule
\end{tabularx}
\caption{The four tasks, their results and the appendices that describe them and their expert reference analyses. A signal strength is a production rate in units of the Standard Model (SM) prediction.}
\label{tab:tasks}
\end{table}

\section{Datasets}
\label{sec:datasets}
\begin{figure}[!b]\centering
\includegraphics[width=\columnwidth]{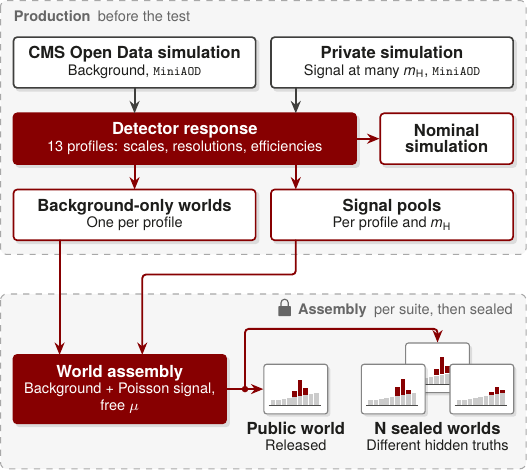}
\caption{Production of the worlds.}
\label{fig:flow}
\end{figure}

The worlds are built from CMS Open Data~\cite{cms_od_dy,cms_od_zz,cms_od_ggzz_4e,cms_od_ggzz_4mu,cms_od_ggzz_2e2mu,cms_od_tt,cms_od_dy_jets,cms_od_dy_ht}, the simulation of proton--proton collisions at $\sqrts=13\TeV$ in the 2016 detector conditions, together with a private signal simulation produced under the same conditions, both in MiniAOD (a reduced analysis format of CMS that keeps all reconstructed particles~\cite{cms_pf_2017}). The Higgs-boson signal is generated with Pythia~8~\cite{pythia82} at varying masses, and its events are then simulated and reconstructed with the official CMS software~\cite{cms_software_2006}, including additional proton--proton interactions within the same or nearby bunch crossings (pileup)~\cite{cms_od_premix}. The detector, its coordinate system and the kinematic variables are described in Ref.~\cite{cms_detector_2008}. For \hwidth, off-shell production through gluon fusion and electroweak processes, with its interference with the $\mathrm{ZZ}$ continuum, is generated with MadGraph5\_aMC@NLO~\cite{madgraph} and passes the same simulation and reconstruction (App.~\ref{app:width}). The \jetcalib worlds are built from simulated Drell--Yan and multijet events of CMS Open Data~\cite{cms_od_dy_nlo,cms_od_qcd_flat,cms_od_qcd_15to30,cms_od_qcd_30to50,cms_od_qcd_50to80,cms_od_qcd_80to120,cms_od_qcd_120to170,cms_od_qcd_170to300,cms_od_qcd_300to470} in a format that keeps all reconstructed particles. The public world of every task is released with the benchmark (see \hyperref[sec:availability]{Code and data availability}), while the sealed worlds used for the results are not. The worlds are indistinguishable from data in their format and carry no generator information. The results of \hfl, \hwidth, \lepcalib and \jetcalib use 40, 30, 10 and 10 sealed worlds of 50, 100, 10 and 1\ifb\footnote{The luminosity (\ifb) is a measure of the number of collision events in a world, and is chosen such that it is sufficient to resolve the differences between the worlds.}, respectively. %

\begin{figure*}[!b]\centering
\includegraphics[width=\textwidth]{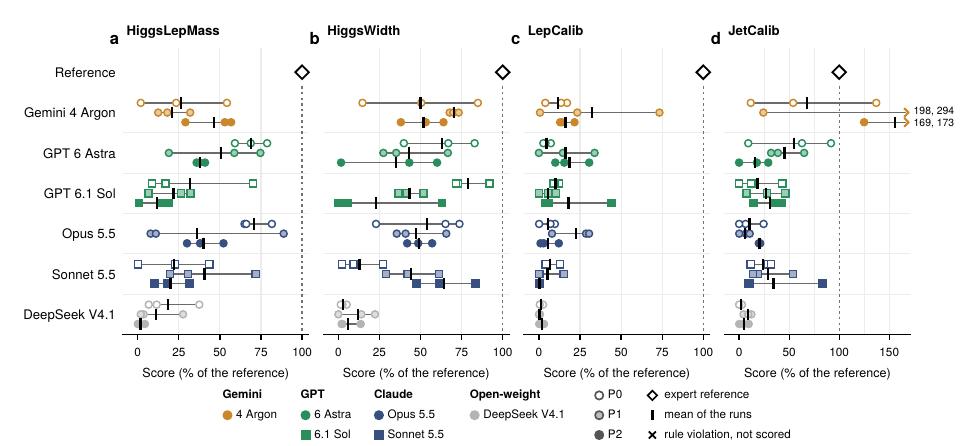}
\caption{Scores of the development sessions as percentages of the expert reference analysis on the same sealed worlds, for the six models with the highest scores (all eleven models in App.~\ref{app:agents}). Within a row the prompt tiers are stacked, P0 at the top, with every run as a point, a line over their range and a vertical bar at their mean. A \jetcalib session beyond the axis is drawn at its edge with its score.} 
\label{fig:results_grid}
\end{figure*}

\subsection{Detector response and signal injection}
\label{sec:modifications}

 The worlds differ from the nominal simulation in the detector response and in the injected signal, and some \hwidth worlds also in the theory prediction. Each world has one of 13 detector profiles, the nominal one or one of twelve modified ones, applied to both its signal and its background. A profile rescales the transverse momentum of every reconstructed particle by a smooth function of $\pt$ and $|\eta|$, smears $\pt$, $\eta$ and $\phi$, and changes the identification and trigger efficiencies by a smooth data-to-simulation ratio. The exact functional forms and their parameters are withheld to prevent cheating. Jets and the missing transverse momentum~\cite{cms_jec_8tev_2017} are rebuilt from the modified particles, so every event stays consistent (App.~\ref{app:datasets}). The signal of a world is generated according to its parameters of interest, the Higgs-boson mass and the signal strength, or in \hwidth the width and the on-shell signal strength, with the production and decay rates, the width and the acceptance changing consistently with the mass. A world may contain no signal.

\subsection{Sealed worlds}
\label{sec:sealed_worlds}

The worlds are produced as shown in Fig.~\ref{fig:flow}. Each world of a task is assembled from a background-only world and a Poisson-distributed number of signal events drawn without replacement from a pool, which holds far more events than any world uses; the worlds are then sealed. The number of background events is Poisson-distributed as well, with a mean set by the luminosity and the cross sections, and its fluctuation exceeds the signal contribution, so the size of a world does not reveal its signal.

The detector profiles and the signal injection act on all leptons, particle-flow candidates~\cite{cms_pf_2017}, jets and the missing transverse momentum alike, so worlds for other analyses or for the calibration of other physics objects can be produced consistently across different background processes and final states.

\section{Results}
\label{sec:results}

\subsection{Tested agents}
\label{sec:performance}

An agent is a language model together with its agent harness, the program that runs the model and connects it to tools. Eleven models of four families were tested, each family under one harness (Table~\ref{tab:models}).
\par\medskip\noindent\begin{minipage}{\columnwidth}\centering\footnotesize
\begin{tabularx}{\columnwidth}{@{}l@{\hspace{0.8em}}>{\raggedright\arraybackslash}X@{\hspace{0.8em}}l@{}}
\toprule
Family & Models & Harness \\
\midrule
Claude & Claude Opus 5.5, Claude Sonnet 5.5 & Claude Code \\
\addlinespace[2pt]
GPT & GPT 6 Astra, GPT 6.1 Sol, GPT 6 Luna & Codex \\
\addlinespace[2pt]
Gemini & Gemini 4 Argon, Gemini 4 Hydrogen,\newline Gemini 3.8 Flash & Antigravity\footnotemark \\
\addlinespace[2pt]
Open-weight & DeepSeek V4.1 Flash, GLM 5.3,\newline Qwen 3.8 Flash Next & OpenCode \\
\bottomrule
\end{tabularx}
\captionsetup{hypcap=false}\captionof{table}{The models tested and the agent harness of each model family.}
\label{tab:models}
\end{minipage}\par\medskip
\footnotetext{Antigravity, unlike Claude Code, Codex and OpenCode, ends a turn after 30 minutes of waiting; a script then sends a fixed message asking the agent to continue.}

Each task has three prompt tiers -- P0, P1, P2. P0 states the task and what to deliver, and P1 and P2 add physics context and an expert plan (App.~\ref{app:prompts}), respectively. The higher tiers stand in for the domain knowledge that an analysis framework such as JFC~\cite{jfc2026} supplies to an agent, and are intended to test how much such guidance adds to the score. Each model was given every prompt tier three times, which gives 99 sessions on each task.  Appendix~\ref{app:agents} shows the scores of all models.

\subsection{Scores}
\label{sec:scores}

Figure~\ref{fig:results_grid} shows the scores. On \hfl, \hwidth and \lepcalib, no session reaches the expert reference, and on \jetcalib six sessions, all of Gemini 4 Argon, score above it. The runs of one model and tier differ by more than the tiers do, so that one session is a poor measure of a configuration. The pipelines that returned no result on any world came mostly from open-weight models. DeepSeek V4.1 Flash is the only open-weight model that reliably delivers working analyses, although 29 of its 36 sessions score below 10\% of the reference. The wall time, the cost and the tokens of the sessions and the evaluation time of the delivered pipelines are compared with the scores in App.~\ref{app:resources}.

\begin{figure}[!h]\centering
\includegraphics[width=\columnwidth]{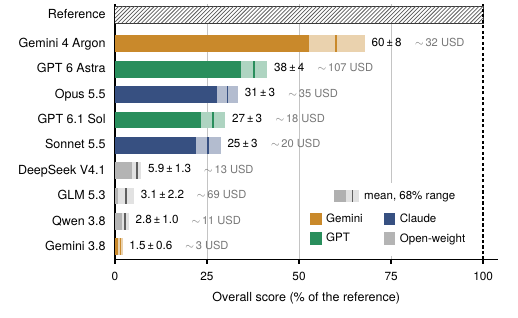}
\caption{Overall score of each model, the mean over the tasks of the mean score of all its sessions on a task, as a percentage of the expert reference. The lighter part of a bar is the 68\% range from a resampling of the sessions and the line the mean. The number in gray is the mean cost of a session at list prices.}
\label{fig:results_overall}
\end{figure}

Figure~\ref{fig:results_overall} combines the tasks into one overall score per model, the mean over the tasks of the mean score of all its sessions on a task. Gemini 4 Argon leads at $60\pm8\%$ of the expert reference, followed by GPT 6 Astra at 38\% and Claude Opus 5.5 at 31\%, while no open-weight model exceeds 10\%. No model leads on every task. The best sessions on \hfl and \hwidth come from Claude Opus 5.5 and GPT 6.1 Sol, respectively, and those on both calibration tasks from Gemini 4 Argon. In general, the scores vary strongly from session to session. For Claude Opus 5.5 on \hfl, the best session reaches 89\% against a mean of 49\%, so the models are limited less by their best result than by how reliably they deliver an analysis pipeline of high quality.

More guidance does not seem to raise the scores. Averaged over the five leading models, the expert plan of P2 brings no gain over P1 on any task, and on \hfl P0 scores highest. The prompt tiers give only basic guidance, and the test of more involved frameworks such as JFC~\cite{jfc2026} is left to future work. Such a test likely requires more tokens per session than standard research subscriptions provide. 

\section{Discussion}
\label{sec:discussion}

\begin{figure}[!b]\centering
\includegraphics[width=\columnwidth]{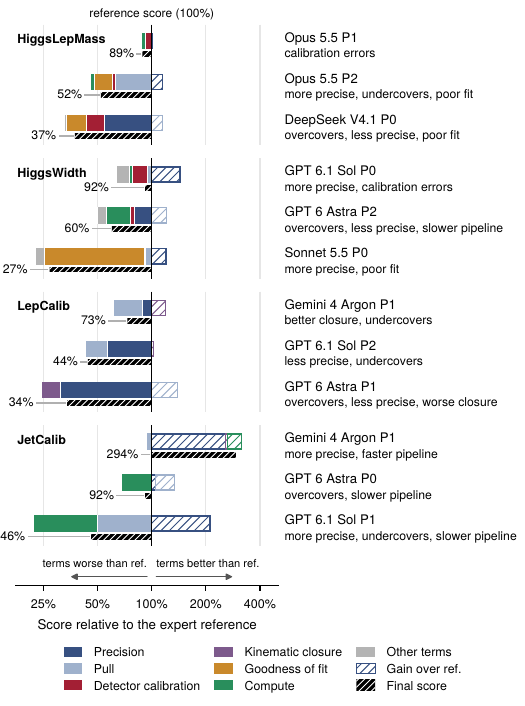}
\caption{Score relative to the expert reference analysis of three selected sessions of each task, by term. Each segment is the factor by which one term changes the score (Eq.~\eqref{eq:score}), lowering it to the left of 100\% and raising it, hatched, to the right. The percentage beside each black hatched bar is the final score.}
\label{fig:results_hero}
\end{figure}

A score $R$ close to 100\% shows that an analysis is as good as the expert reference analysis across the sealed worlds.
Such an analysis recovers the injected values with accurate uncertainties, calibrates its detector and describes the data with its model. A high score supports trust in an analysis under the tested conditions and complements the review of an analysis in a collaboration.
The individual terms locate the shortfalls of an analysis. By Eq.~\eqref{eq:score}, the score relative to the reference is a product of one factor per term. Figure~\ref{fig:results_hero} shows the segments for three selected sessions of each task.

On \hfl, the leading session of Claude Opus 5.5 loses little, mostly in the detector calibration, while its session at P2 undercovers and DeepSeek V4.1 Flash is less precise, with a model that describes the pseudodata poorly. The other \hfl sessions that score well likewise lose mainly in the detector calibration. Those that score poorly often fail the goodness-of-fit test as well, so that this test identifies them without the truth. 

On \lepcalib, nearly all of the loss comes from the accuracy of the corrections themselves, and the quality terms otherwise find no issues. In the kinematic closure, the agreement of the corrected simulation with the pseudodata at the $\mathrm{Z}$ peak, the leading Gemini 4 Argon and GPT 6.1 Sol sessions are better than the reference. %
On \jetcalib, the sessions that score above the reference use varying methods, some fitting all corrections together and others following the factorized steps of the CMS calibration~\cite{cms_jec_8tev_2017}. Their gains are in jet energy scale, where their corrections are about as close to the truth as those of the reference but their uncertainties are about half as large. The uncertainties of the reference are dominated by its modeling choices and exceed the variations within the worlds, which leads to lower score. Future worlds produced with alternative generators would add the modeling differences between data and simulation that such uncertainties protect against.
 GPT 6.1 Sol at P1 loses in pipeline wall time and in the pull piece more than it gains in precision.
The terms of the best session of each model and tier are shown in App.~\ref{app:loss}.

Figure~\ref{fig:results_plane} shows the precision and pull pieces of the measurement term for the sessions that reach a tenth of the reference. The sessions undercover, with a pull piece above the reference's, far more often than they overcover. Many sessions are more precise than the reference, in particular on \jetcalib, but there their larger pulls cost on average more than the precision gains. The agents' uncertainties are thus often as small as the reference's, but less often accurate.

The comparison with the injected truths over many sealed worlds makes these observations possible. An agent can deliver a pipeline that runs and reports small uncertainties while its results lie further from the truth than these uncertainties allow, which this comparison shows directly. It also establishes when an analysis is better than the expert reference, which thus serves as a common standard, not an upper limit.

The worlds differ from the nominal simulation mainly in the injected detector response and signal, and worlds with alternative generators would extend the test to the physics modeling. The score depends on the expert reference analysis and on the chosen weights. Future work can vary both, with further references and with the stored terms of every evaluation. All worlds come from the simulation of one experiment, and the extension to other experiments described below would test whether the conclusions hold there.

\section{Conclusions and outlook}
\label{sec:conclusion}

We have presented the building blocks of \bench, a benchmark for autonomous physics analysis. Anchoring the score to the expert reference analysis and to standard statistical tools provides an objective way of evaluating correctness, without expert review of every analysis and without a language-model judge. The score separates accurate analyses from overconfident, biased or failing ones. As agents take on more of the analysis work, we argue that such quantitative and repeatable tests should become common practice in their evaluation.

\bench can be extended to further kinds of analysis, such as calibrations of other physics objects and frontier measurements that need multivariate discriminants. Future tasks could also include other experiments, such as ATLAS, LHCb, neutrino detectors and gravitational waves.

Building such tasks remains the work of human experts. Which analyses become tasks, which properties of the worlds vary and how strongly, and what a correct analysis has to deliver are decisions of physics judgment, made once per task and then applied to every agent.

\section*{Code and data availability}
\phantomsection\label{sec:availability}
The benchmark is hosted at \url{https://hep-ex-bench.github.io}, the entry point to its datasets, its harness and the scoring of new pipelines. The public worlds and the nominal simulation are released on Hugging Face at \url{https://huggingface.co/hep-ex-bench} and archived on Zenodo~\cite{heb_data_hfl,heb_data_hwidth,heb_data_lepcalib}. They are built from CMS simulation, the public samples of CMS Open Data~\cite{cms_od_dy,cms_od_zz,cms_od_ggzz_4e,cms_od_ggzz_4mu,cms_od_ggzz_2e2mu,cms_od_tt,cms_od_premix} and private samples with the same detector simulation. The harness that runs pipelines and scores them is open source at \url{https://github.com/hep-ex-bench/hep-ex-bench}. The sealed worlds and the code that builds them are withheld by design. A pipeline developed on the public world can be scored on the sealed worlds of this paper on request to the authors. 

\section*{Acknowledgments}
\begin{figure*}[!t]\centering
\includegraphics[width=\textwidth]{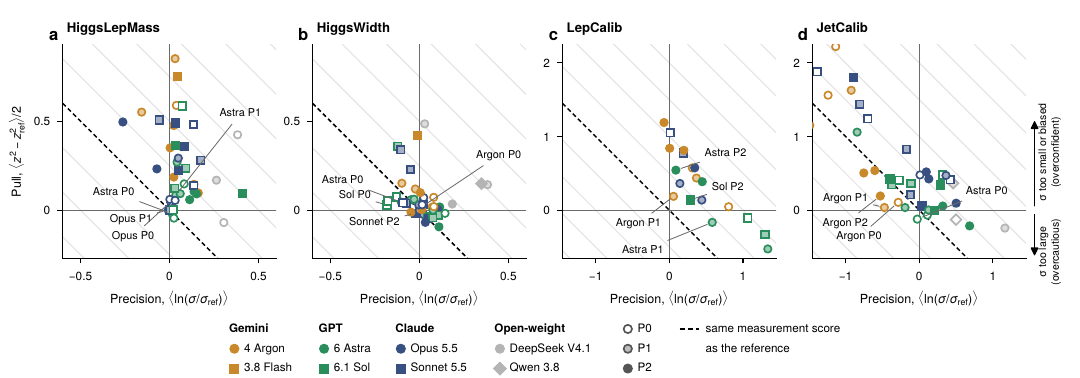}
\caption{The precision and pull pieces of the measurement term, Eq.~\eqref{eq:decomp}, of every session of the models of Fig.~\ref{fig:results_grid} scoring at least 10\% of the expert reference analysis, both relative to the reference, with $\pull$ the pull. On the dashed line the measurement term equals the reference's, below it the session measures better, and the gray lines mark equal measurement terms. A bias and an understated uncertainty both raise the pull piece; only an overstated uncertainty lowers it.}
\label{fig:results_plane}
\end{figure*}
We thank CERN and the CMS Collaboration for the open data and simulation that this work is built on; neither CMS nor CERN endorses the results. This work is supported by the National Science Foundation under Cooperative Agreement PHY-2019786 (The NSF AI Institute for Artificial Intelligence and Fundamental Interactions, \url{http://iaifi.org/}). A.~N., E.~M., and P.~H. are supported by the NSF-funded A3D3 Institute (NSF-PHY-2117997), DOE Early Career award FY2021, ``Harnessing the Large Hadron Collider with New Insights in Real-Time Data Processing and Artificial Intelligence'', and DOE Genesis award ``Agentic AI for Real-time Expedited Discovery from High Complexity EIC Data Streams". D.~G. is supported by the Swiss National Science Foundation Postdoc.Mobility Fellowship n 235361. V.~B.~P. acknowledges financial support from the Real Colegio Complutense at Harvard University Postdoctoral Research Fellowship. MK and JK are supported by the US Department of Energy (DOE) under Grant No.~DE-AC02-76SF00515. LH is supported by the Excellence Cluster ORIGINS, which is funded by the Deutsche Forschungsgemeinschaft (DFG, German Research Foundation) under Germany’s Excellence Strategy - EXC-2094-390783311 and BMFTR Project SciFM 05D25WO2. This work used resources available through the National Research Platform (NRP) at the University of California, San Diego. NRP has been developed, and is supported in part, by funding from National Science Foundation, from awards 1730158, 1540112, 1541349, 1826967, 2112167, 2100237, and 2120019, as well as additional funding from community partners. The computations in this paper were also run on the FASRC Cannon cluster supported by the FAS Division of Science Research Computing Group at Harvard University. We thank Anthropic, Google and OpenAI for the free or discounted access to their models through their research programs, which made the development sessions of this work possible.

\makeatletter
\def\bib@stripclass#1 [#2]#3\@nil{#1}
\newcommand{\Eprint}[2]{\href{#1}{\bib@stripclass#2 []\@nil}\ignorespaces}
\immediate\write\@auxout{\string\citation{apsrev42Control}}
\makeatother
\renewcommand{\bibfont}{\footnotesize\renewcommand{\_}{\textunderscore\allowbreak}}
\setlength{\bibsep}{2pt plus 1pt}
\bibliographystyle{apsrev4-2}
\newpage %
\bibliography{references}

\clearpage
\appendix
\counterwithin{figure}{section}\counterwithin{table}{section}
\renewcommand{\thefigure}{\thesection.\arabic{figure}}\renewcommand{\thetable}{\thesection.\arabic{table}}
\makeatletter\@ifundefined{theHfigure}{}{\renewcommand{\theHfigure}{\thesection.\arabic{figure}}\renewcommand{\theHtable}{\thesection.\arabic{table}}}\makeatother
\raggedbottom
\section{Extended results}
\label{app:results}

\subsection{Scores of the development sessions}
\label{app:agents}
Figure~\ref{fig:results_ladder_all} shows the scores of Fig.~\ref{fig:results_grid} for all models, one panel per task.

\begin{figure*}[!htbp]\centering
\includegraphics[width=\textwidth]{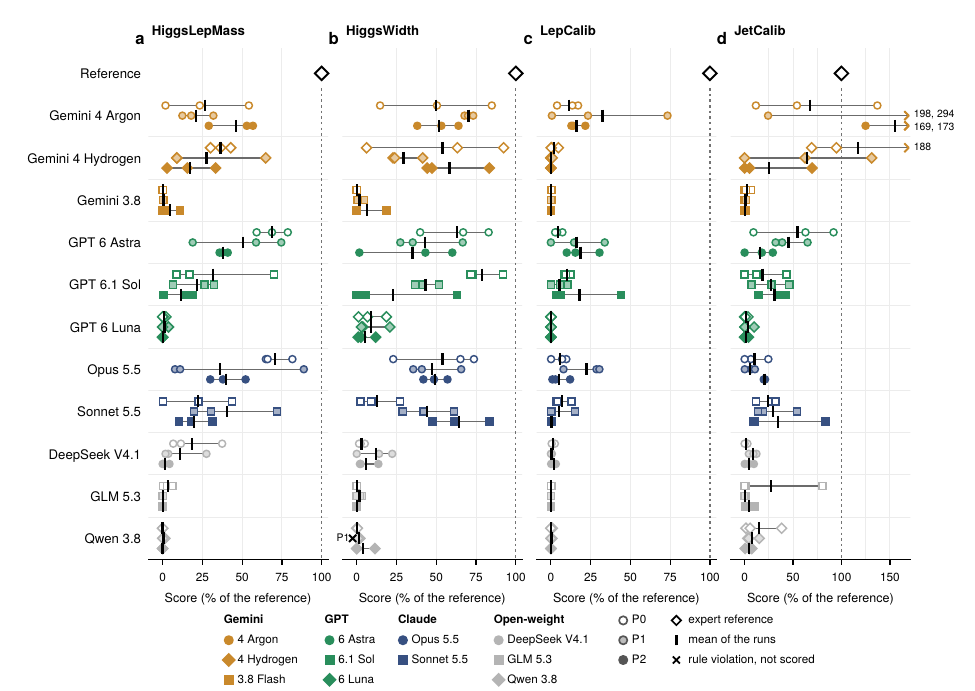}
\caption{Scores of the development sessions of all models, one panel per task, as in Fig.~\ref{fig:results_grid}. GPT~6 Luna and Gemini~4 Hydrogen, which are in no other figure, are included.} %
\label{fig:results_ladder_all}
\end{figure*}

\subsection{Reasoning effort}
\label{app:effort}

Every session of the campaign runs at the default reasoning effort of its agent harness. To measure the dependence of the score on this setting, additional sessions of Claude Opus~5.5 were run on \hfl with the P1 prompt at the low, high and max levels of the reasoning effort of Claude Code, with three sessions each. All other settings are the same.

Figure~\ref{fig:effort} shows the scores of three sessions for each of the four effort levels. At low effort, all three sessions score at most 11\% of the expert reference analysis, using about a fifth of the tokens of a default session. From medium upward, the sessions of one level differ by more than the levels do. The three default sessions range from 8\% to 89\%, and the mean scores at medium, high and max are 36\%, 49\% and 38\%. With three sessions per level, no difference between these three levels is resolved. The sessions at max effort are the most expensive, at 69~USD on average at list prices against 34~USD at the default, and do not score higher.

\begin{center}
\includegraphics[width=\colfigwidth]{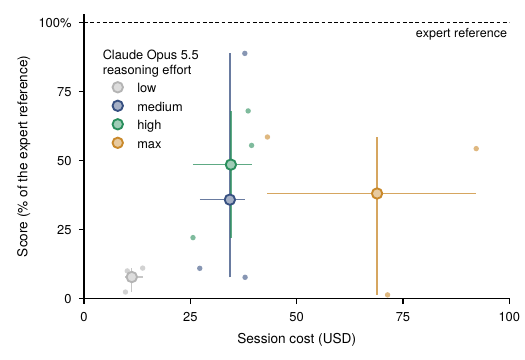}
\captionsetup{hypcap=false}\captionof{figure}{Claude Opus~5.5 scores on \hfl with the P1 prompt at four levels of reasoning effort, with the mean score of each level against the mean cost of its development sessions at list prices. The error bars span the range of sessions at that level. Individual trials are shown as small points in the color of the level.}
\label{fig:effort}
\end{center}

\subsection{Pulls and loss terms of the development sessions}
\label{app:loss}
Figure~\ref{fig:results_pulls_hfl} shows the measurement term pulls of the best \hfl session of each model and tier. Figures~\ref{fig:results_terms_hfl}, \ref{fig:results_terms_hwidth}, \ref{fig:results_terms_lepcalib} and~\ref{fig:results_terms_jetcalib} show all terms of the total loss of the best session of each model and tier, for each task on its own. Each row stacks the weighted losses of the terms relative to the expert reference analysis, $w_k(\Delta_k-\Delta_k^{\mathrm{ref}})$, preserving order. The considered terms are the precision and pull pieces of the measurement term of Eq.~\eqref{eq:pieces}, the detector-calibration term in \hfl and \hwidth or the kinematic-closure term in \lepcalib and \jetcalib, goodness of fit, compute, and the sum of the completeness and truthfulness terms. The precision and pull terms also include contributions from world and quantity combinations and from failed cases. The segments add up to $\Dtot-\Dtot^{\mathrm{ref}}$. The axis is the score relative to the reference on a logarithmic scale, on which each segment is the factor of one term in Eq.~\eqref{eq:score}, as in Fig.~\ref{fig:results_hero}. The hatched black bars under each stack represent the total session score over the reference. A row with a loss beyond the axis is cut with a break mark. Sessions scoring below 10\% are not shown.

\subsection{Resources of the sessions}
\label{app:resources}

Figure~\ref{fig:results_resources} relates the score to the wall time of the development session, the inference time and its cost. We also indicate the number of tokens of the best performing sessions. On \hfl, the 16 sessions that score above half of the reference use between 2 and 7.5\,h of the available 7.5\,h and cost 14 to 176~USD. For most models neither longer sessions nor higher cost yields a higher score. Over the four tasks, the score rises only weakly with the cost of a session and hardly with its wall time. The median session of Gemini 4 Argon, Claude Sonnet 5.5 and Claude Opus 5.5 ends after 3 to 4\,h, while the sessions of GPT 6 Astra and GPT 6.1 Sol use nearly all of the 7.5\,h. %
The best \hwidth session uses 7.5\,h and costs 23~USD, the best \lepcalib session 2.7\,h and 34~USD, and the best \jetcalib session 3.2\,h and 41~USD. %

Concerning the evaluation time, for \hfl the pipelines take between 0.1 and 7.9 times the time of the reference, narrowing to 0.3–3.0 times for sessions scoring above 50\% of the reference. %
On \hwidth, the pipelines take between 0.3 and 8.5 times the time of the reference. On \lepcalib, most pipelines run faster than the reference, and the time does not separate the sessions either. On \jetcalib, the pipelines take between 0.1 and 28 times the time of the reference, and most run slower than it. Therefore, evaluation time does not determine session performance.

\begin{figure}[!htbp]\centering
\includegraphics[width=\colfigwidth]{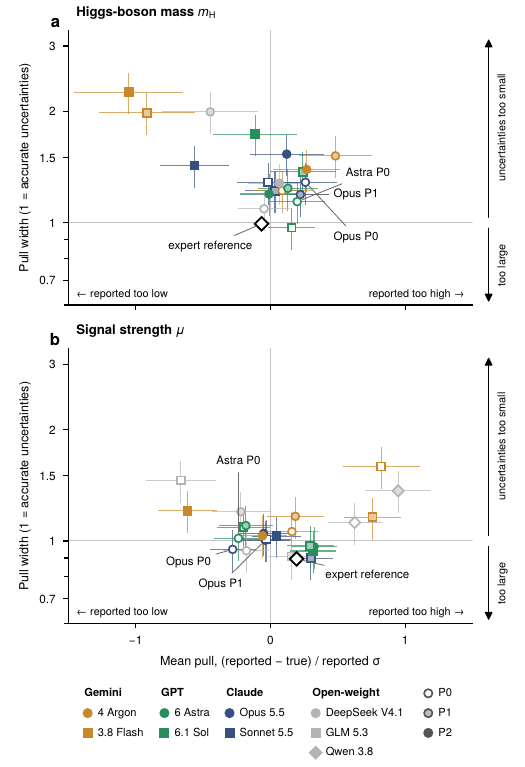}
\caption{Mean pull and pull width of the reported $\mh$ (a) and $\mu$ (b) of the best \hfl session of each model and tier over its sealed worlds, with $\pm1$ standard error from the number of worlds. An unbiased analysis with accurate uncertainties sits at (0, 1), and the expert reference analysis are marked with open diamonds. The three sessions with the highest scores are labeled. Sessions outside the shown range are left out.}
\label{fig:results_pulls_hfl}
\end{figure}

\begin{figure*}[!htbp]\centering
\includegraphics[width=\textwidth]{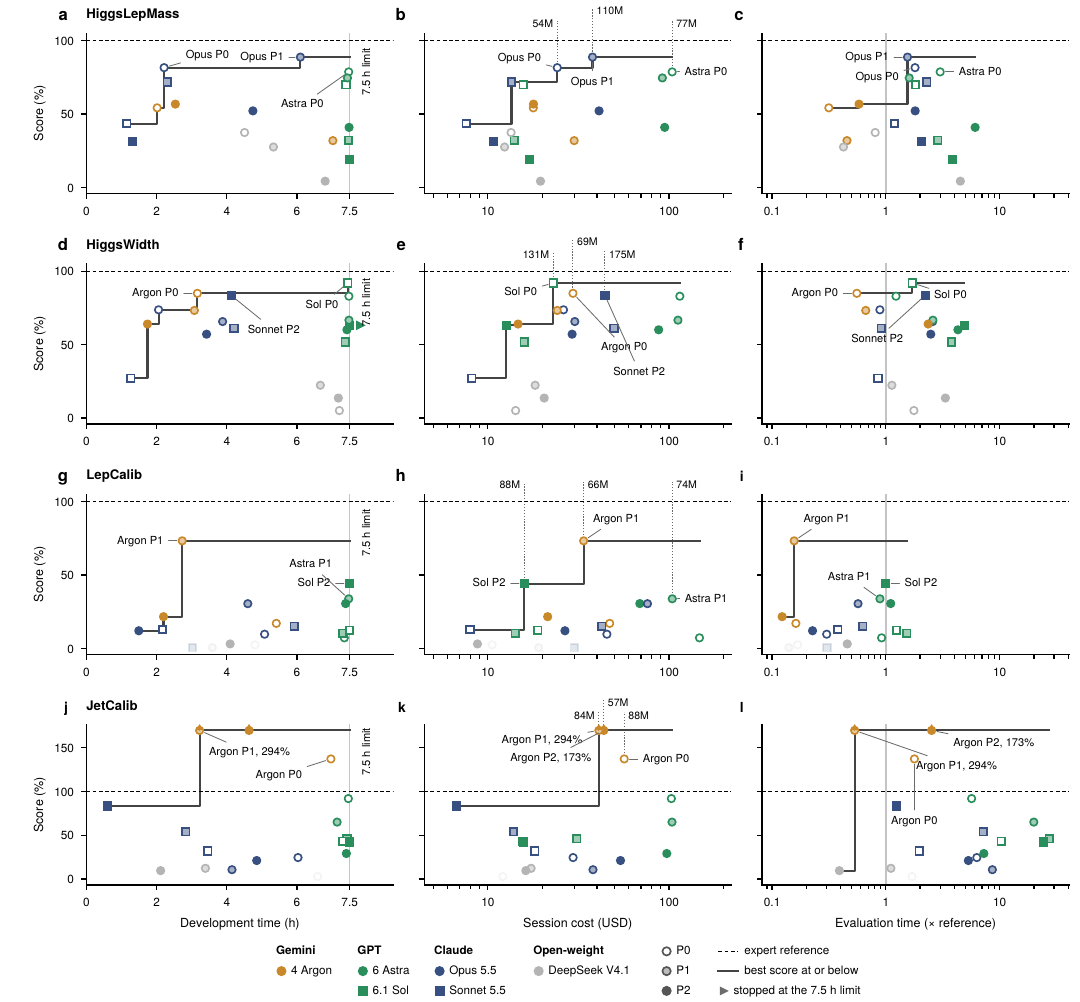}
\caption{Scores for the best sessions of the models in and tiers against the wall time of the development session (left), their cost at list prices (middle) and the evaluation time of the delivered pipeline per world relative to the expert reference analysis (right). We only show the models presented in Fig.~\ref{fig:results_grid}. Sessions at or below 3\% are drawn faint, and only sessions with a score above zero are shown. The numbers above the cost panels are the tokens used by the named sessions, in millions.}
\label{fig:results_resources}
\end{figure*}

\begin{figure*}[!p]\centering
\begin{minipage}[t]{0.485\textwidth}\centering\vspace{0pt}%
\includegraphics[width=0.96\linewidth]{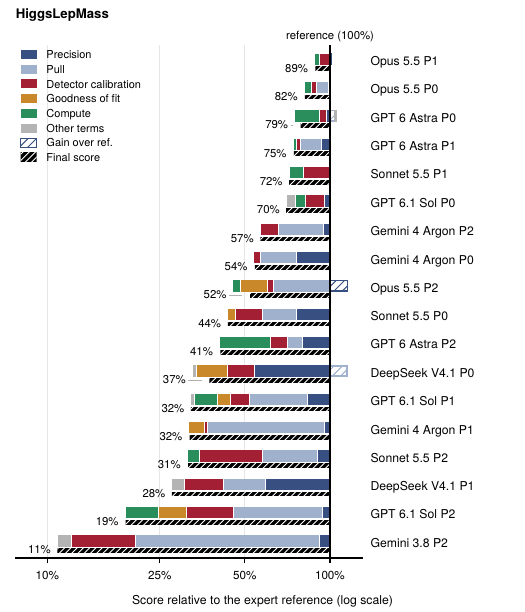}
\caption{Score relative to the expert reference analysis of the best \hfl session of each model and tier, with the same term decomposition as Fig.~\ref{fig:results_hero}.}
\label{fig:results_terms_hfl}

\vspace{4pt}
\includegraphics[width=0.96\linewidth]{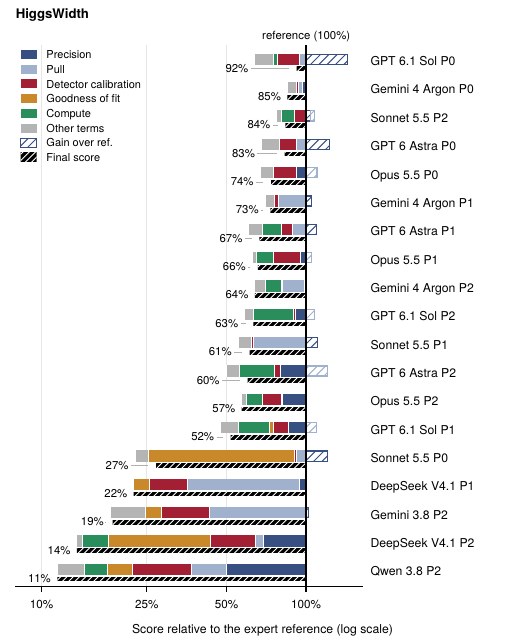}
\caption{Score relative to the expert reference analysis of the best \hwidth session of each model and tier, with the same term decomposition as Fig.~\ref{fig:results_hero}.}
\label{fig:results_terms_hwidth}

\end{minipage}\hfill
\begin{minipage}[t]{0.485\textwidth}\centering\vspace{0pt}%
\includegraphics[width=0.96\linewidth]{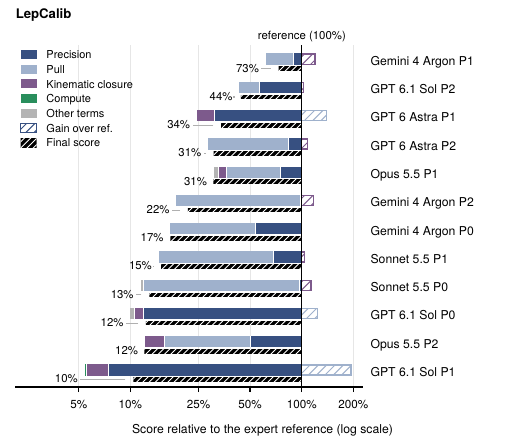}
\caption{Score relative to the expert reference analysis of the best \lepcalib session of each model and tier, with the same term decomposition as Fig.~\ref{fig:results_hero}.}
\label{fig:results_terms_lepcalib}

\vspace{4pt}
\includegraphics[width=0.96\linewidth]{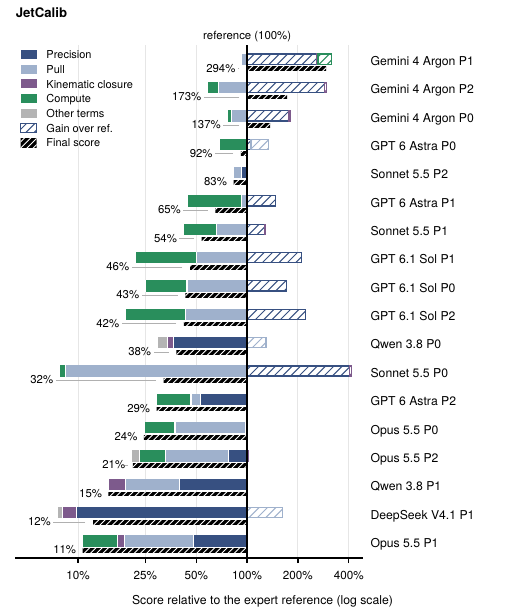}
\caption{Score relative to the expert reference analysis of the best \jetcalib session of each model and tier, with the same term decomposition as Fig.~\ref{fig:results_hero}.}
\label{fig:results_terms_jetcalib}
\end{minipage}
\end{figure*}

\clearpage
\section{Scoring details}
\label{app:metrics}

This appendix describes the score in detail and documents tests on its robustness. We consider two types of tasks: the \hfl and \hwidth fit measurements, where the score evaluates the parameters of a statistical model, and the \lepcalib and \jetcalib calibration tasks, where the score evaluates correction functions. Full documentation is available in the agentic harness documentation~\cite{sbh_repo}.

\subsection{Likelihood scoring}
\label{app:likelihood}
The penalty is the negative logarithm of the density that the analysis assigns to the truth, whatever the shape of that density. This is the logarithmic score, a \emph{strictly proper} scoring rule~\cite{good_1952,gneiting_raftery_2007}. The pipeline of a fit measurement delivers its statistical model with its report. The likelihood of a grid of fixed masses is also provided in the case of \hfl. The reported values are not used directly used to compute the measurement term. Instead, they are obtained by refitting the delivered likelihood.

The signal strength $\mu$ is scored as a density. At each grid mass, within the evaluation range stated in the contract, the likelihood is profiled as a function of $\mu$. The profiles are summed with the weights of a fixed prior on the mass into a density $p(\mu)$. Considering the injected value $t$ and the density of the reference on the same world $p_{\mathrm{ref}}$, the penalty term for each world is
\begin{equation}
  \Delta_i = -\ln p(t) - H[p_{\mathrm{ref}}] ,\qquad H[p]=-\int p(\mu)\ln p(\mu)\,\mathrm{d}\mu ,
  \label{eq:case_loss}
\end{equation}
which reduces to Eq.~\eqref{eq:pieces} when both densities are Gaussian. Its precision piece is $H[p]-H[p_{\mathrm{ref}}]$ and its pull piece $-\ln p(t)-H[p]$. The combination of the worlds in Eq.~\eqref{eq:bias} needs an uncertainty and a pull for each world. For a density that is not Gaussian, $\sigma_i$ is the width of the Gaussian with the same entropy, $\sigma_i=\mathrm{e}^{H[p]}/\sqrt{2\pi\mathrm{e}}$, and the pull is $\pull_i=\operatorname{sign}(\hat\mu-t)\sqrt{2\ln\frac{p(\hat\mu)}{p(t)}}$, with $\hat\mu$ the mode of $p$. For a Gaussian density both reduce to those of Eq.~\eqref{eq:pieces}. The worlds of \hfl may contain no signal.

The mass is computed from the delivered grid. With $q(m)=-2\ln\frac{L(m)}{L_{\max}}$ for the profile likelihood $L(m)$ at the fixed mass $m$ and its largest value $L_{\max}$ on the grid, the value $x_i$ is the position of the minimum of the parabola through the three grid points around the smallest $q$, and the uncertainty $\sigma_i$ is half the width of the interval with $q<1$. The width ratio $\xi$ of \hwidth is scored in the same way by using the density $p(\xi)\propto\mathrm{e}^{-q(\xi)/2}$ obtain from the evaluator scan of the profile likelihood of the delivered model over the range of 0 to 100 stated in the contract. Its on-shell signal strength is read as a Gaussian with the uncertainty of its interval on each side.

\subsection{Calibration scoring}
\label{app:calibration}
We score the calibration over kinematic regions defined by $a=1,\dots,M$. At point $a$ of world $i$, we deal with the reported value $x_{ia}$, the uncertainty $\sigma_{ia}$, to which a small minimum set by the task is added in quadrature, the injected truth $t_{ia}$ and the pull $\pull_{ia}=(x_{ia}-t_{ia})/\sigma_{ia}$. The variables in Eqs.~\eqref{eq:pieces} and~\eqref{eq:bias} are constructed as follows:
\begin{itemize}[nosep, leftmargin=1.4em]
\item $\sigma_i$, the geometric mean of the $\sigma_{ia}$ over the points,
\item $\sigma_{\mathrm{ref},i}$, the same geometric mean for the reference on that world, multiplied by the root-mean-square pull of the reference over the points and cases of the group of corrections defined below,
\item $\pull_i^2$, the squared sum of the mean of the $\pull_{ia}$ and the standard deviation of the $\pull_{ia}$ over the points, which we call $s_i$.
\end{itemize}
 The combination of the worlds in this case also takes into account the shared biases over the worlds, resulting in a final score,
\begin{equation}
  \Delta_p = \frac1N\sum_{i=1}^{N}\Delta_i + \frac{\zstar^2+\sstar^2-1}{2N} ,\qquad \sstar^2=\sigstar^2\sum_{i}\frac{s_i^2}{\sigma_i^2} ,
  \label{eq:spread}
\end{equation}
This expression reduces to Eq.~\eqref{eq:bias} for a single reported value, i.e., $s_i=0$.

In \lepcalib, the analysis produces a group of four corrections: the scales, the resolutions, the identification efficiencies or the trigger efficiencies. There are 2 groups for \jetcalib: the scale and the resolution, each in the central and endcap regions of the detector. Each requested correction has its penalty term quantified by Eq.~\eqref{eq:spread} for every flavor. The penalties are averaged within a group, and the measurement factor is the mean over the $n$ groups.

\subsection{Analyses without a measurement}
\label{app:missing}
An analysis has no measurement when the run on its world fails under the evaluation contract, crashes, hits a time or memory limit, produces missing or invalid outputs, or when its outputs do not determine the quantity. The penalty in these cases is as if the nominal value $x_0$ and the prior width $\sigma_0$ of the task had been reported,
\begin{equation}
  \Delta_i = \max\big(F_i,0\big) ,\quad
  F_i = \underbrace{\ln\frac{\sigma_0}{\sigma_{\mathrm{ref},i}}}_{\text{precision}} + \underbrace{\tfrac12\big(\tfrac{(x_0-t_i)^2}{\sigma_0^2} - 1\big)}_{\text{pull}} .
  \label{eq:case}
\end{equation}
The nominal value and the prior width are fixed by the task. For the measurements in \hfl and \hwidth, $x_0$ corresponds to the center of their stated ranges. For a \lepcalib or \jetcalib correction, $x_0$ is the nominal value stated in the request and $\sigma_0$ the root-mean-square deviation of the injected truths from it.

\subsection{Several quantities}
\label{app:quantities}
 When dealing with $n$ measurements, each has its own penalty $\Delta_p$ and the scores are averaged,
\begin{equation}
  \Sv = \underbrace{\frac1n\sum_{p=1}^{n}\mathrm{e}^{-\gamma\,\Delta_p}}_{\text{measurement}} \;\times\; \underbrace{\mathrm{e}^{-\gamma\sum_k w_k\Delta_k}\vphantom{\sum_{p=1}^{n}}}_{\text{quality}} ,
  \label{eq:quantities}
\end{equation}
where the first factor represents $\Dphys$ and reduces to Eq.~\eqref{eq:score} for $n=1$.

\subsection{Additional quality terms}
\label{app:quality}

For the detector calibration, each detector effect $j$ the analysis logs, with uncertainty $s_j$ and pull $\pull_j$ against the injected effect, is scored like a measurement, with the task's target precision $\tau_j$ in place of the reference,
\begin{equation}
  \Delta_j = \min\Big(\max\Big(\ln\frac{s_j}{\tau_j},0\Big) + \tfrac12\,(\pull_j^2-1),\; \max(U_j,4)\Big) ,
  \label{eq:cal}
\end{equation}
and $\Dcal$ is the average over effects and worlds. An unlogged effect is penalized by $U_j$, the penalty of Eq.~\eqref{eq:case} with the prior of the effect and $\tau_j$. The cap keeps one effect from outweighing the measurement, and an uncertainty below the target does not contribute.

For the goodness of fit, the produced model is tested on every world by a saturated-likelihood test~\cite{baker_cousins_1984}, and the $p$-values of the worlds that give one, each no smaller than a floor of 0.005, are combined by Fisher's method~\cite{fisher_1925} into one $p$-value with Gaussian significance $Z(p)$. The penalty is the significance beyond that of $p=0.01$, one unit per standard deviation,
\begin{equation}
  \Dgof = \min\big(\max\big(0,\,Z(p)-Z(0.01)\big),\,4.3\big) ,
  \label{eq:gof}
\end{equation}
 so a model that describes the pseudodata has no penalty. With the minimum floor, a poor fit on a single world does not impact the result. A world whose model gives no $p$-value is left out of the combination and counted at the cap, and the term is the mean over all worlds of the penalty of Eq.~\eqref{eq:gof} and the cap.

 The other terms of Table~\ref{tab:quality} are as follows, each averaged over the worlds. A term is zero when its check passes, and the kinematic closure and the compute term are negative for a pipeline that does better than the reference. Tolerances and caps are fixed per task in the harness~\cite{sbh_repo}.
\begin{itemize}[nosep, leftmargin=1.4em]
\item Truthfulness, $\Dtruth$. The reported values and uncertainties are compared with the evaluator's refit of the produced model. The penalty is the difference of the values in units of the refit uncertainty, plus the logarithm of the factor by which the reported uncertainty falls below the refit one.
\item Kinematic closure, $\Dclosure$. The reported corrections are applied to the simulation, and its $\Zll$ mass distribution is scored against the pseudodata bin by bin as in Eq.~\eqref{eq:pieces}, in categories of the dilepton kinematics and with the uncertainties of the corrections propagated. In \jetcalib, the distributions are the $\pt$ balance of $\mathrm{Z}+\mathrm{jet}$ events and the asymmetry of dijet events. The penalty is the difference from the same quantity for the reference.
\item Completeness, $\Dsyst$. The fraction of the standard systematic uncertainties of the measurement that the produced model, or for \lepcalib the report, does not account for.
\item Compute, $\Dcomp$. The base-2 logarithm of the ratio of the wall time of the pipeline to that of the reference, beyond a tolerance of 1.5. A faster pipeline gains at most one unit, and only while the rest of $\Dtot$ is close to zero. The wall time of a pipeline and of the reference is the median over the timed worlds. Every world is timed in \hfl, \hwidth and \lepcalib. In \jetcalib, ten of the 32 worlds are timed, on one type of compute node, and all other terms use the 32 worlds.
\end{itemize}

\subsection{Bootstrap uncertainty of the score}
\label{app:bands}
The band of $\Sv$, of $R$ and of each term is a paired bootstrap over the worlds. The sealed worlds of a task are resampled with replacement 1000 times, in blocks of the worlds that share a detector profile, and the whole score is recomputed on every resample. The band is the 16th to 84th percentile over the resamples, widened for the small number of blocks $n_{\mathrm{b}}$ by $\sqrt{\frac{n_{\mathrm{b}}}{n_{\mathrm{b}}-1}}\,\frac{t_{n_{\mathrm{b}}-1}(0.84)}{z(0.84)}$, the correction of the bootstrap variance for the number of blocks times the ratio of the 84\% quantiles of Student's $t$ distribution with $n_{\mathrm{b}}-1$ degrees of freedom and of the unit Gaussian, a factor of $1.22$ at six blocks. The band of $R$ is that of the ratio of the scores of the analysis and of the reference on the same resamples.

The bands are also quoted on the two pieces of the measurement term. The figures of Secs.~\ref{sec:results} and~\ref{sec:discussion} show the pieces of Eq.~\eqref{eq:pieces} averaged over the $N_{\mathrm{m}}$ worlds of one quantity with a measurement, each as its difference from that of the reference at unit pull,
\begin{equation}
  \langle\Delta\rangle = \underbrace{\frac{1}{N_{\mathrm{m}}}\sum_i\ln\frac{\sigma_i}{\sigma_{\mathrm{ref},i}}}_{\text{precision}} + \underbrace{\frac{r^2-1}{2}\vphantom{\sum_i\frac{\sigma_i}{\sigma_{\mathrm{ref},i}}}}_{\text{pull}} ,\qquad r^2=\frac{1}{N_{\mathrm{m}}}\sum_i\pull_i^2 ,
  \label{eq:decomp}
\end{equation}
with $r$ the root-mean-square pull. The pull term is zero for $r=1$, positive for $r>1$ and never below $-\tfrac12$.

\begin{figure}[!t]\centering
\includegraphics[width=\colfigwidth]{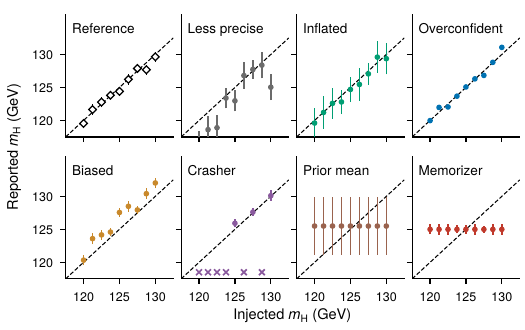}
\caption{Archetypes of an analysis in a toy with one parameter, $\mh$: reported against injected values on nine worlds, with enlarged uncertainties for visibility. Crosses mark the worlds on which Crasher returns no result.}
\label{fig:toy_reported}
\end{figure}

\begin{figure}[!t]\centering
\includegraphics[width=\colfigwidth]{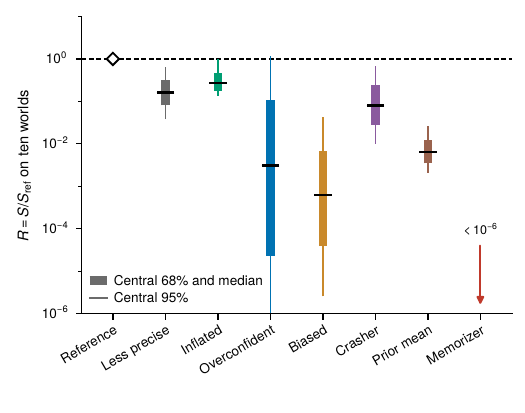}
\caption{Ratio $R$ of the score of each archetype of Fig.~\ref{fig:toy_reported} to that of the expert reference analysis on the same worlds, over 2000 sets of ten sealed worlds. The score of the toy contains only the measurement term, $\mathrm{e}^{-2\Dphys}$. Memorizer lies far below the range shown.}
\label{fig:toy_worlds}
\end{figure}

\subsection{Robustness to failure modes and cheating}
\label{app:adversarial}
The score is designed so that neither a failure mode of an analysis nor an attempt to cheat is capable of increasing scores above the expert reference analysis. The design is tested with failures injected into the real tasks and with a toy study of the number of sealed worlds that the separation requires.

\label{app:adversarial_tests}
To test whether the evaluation is sound, we built nine flawed submissions and ten hostile pipelines, each of which deliberately errs or cheats in one way. The hostile pipelines check the isolation of the harness~\cite{sbh_repo}. They try to read the truth, open the network, escape through symbolic links, write outside the run, or exhaust memory or time. Others flood the log, return malformed results, behave non-deterministically or ship cached data.

Most flawed submissions are the reference with one behavior changed. They ignore the pseudodata, fail on part of the worlds, overstate or understate every uncertainty, or shift every reported value by a multiple of its uncertainty. Others fix a parameter at a remembered value, skip the in-situ calibration, report a measurement that the pseudodata contradict, or deliver their result through a symbolic link. The score penalizes each of them through the term that it affects.
On \hfl, a submission that overstates its uncertainties threefold reaches 88\% of the reference, one that ignores part of the pseudodata 76\%, one that understates its uncertainties 34\%, one without lepton calibration 4\%, and the two that report a memorized value and a constant less than 1\% each. A pipeline that crashes on more than half of the worlds is penalized for each of them and reaches 1\%. On \lepcalib, a crashing pipeline fails the run and scores 0, a submission with fabricated probes reaches 69\% of the reference, and every other one stays below 27\%. One example of a typical failure within \lepcalib occurs for a submission that declares a fifth of the failing identification probes as passing has large pulls only in the manipulated corrections, so the score locates them. The threats that these tests address are listed in Sec.~\ref{sec:multiworld}. %

\subsection{Dependence of the score on the number of sealed worlds}
\label{app:toy}
 The number of sealed worlds decides how reliably such failures are caught. A task is scored on many sealed worlds because the score on one world does not separate an analysis that measures from one that returns the expected value. How the separation grows with the number of worlds is studied in a toy with one parameter, $\mh$. The toy is constructed through world assembly where the expert reference analysis reports the injected value with a Gaussian fluctuation of its uncertainty $\sref$. Seven failure examples following App.~\ref{app:adversarial_tests} are considered, each mimicking one way in which an analysis goes wrong (Fig.~\ref{fig:toy_reported}). These failures consist of cases where the result: 
\begin{itemize}[nosep, leftmargin=1.4em]
\item \emph{ is less precise} has accurate uncertainties 2.5 times larger than those of the reference, as an analysis with a poorer resolution or a smaller selected sample has,
\item \emph{ produces inflated uncertainties} with correct central values, but with three times its uncertainty,
\item \emph{is overconfident} with an analysis that yield a third of the reference uncertainty, through the incorrect exclusion of systematic uncertainties,
\item \emph{ is biased} with a shifted of $2\,\sref$ on every world, reflecting an analysis with an uncorrected offset,
\item \emph{ that crashes} resulting in results that do not converge on half of the worlds,
\item \emph{performs no work} with a prior mean, center and width of the initial range stated, but with no measurement, and,
\item \emph{memorizes} reports at 125\GeV with the uncertainty of the reference on every world and applies it generically to all analyses.
\end{itemize}
Scored on a single world at the SM value, Memorizer scores higher than the expert reference analysis. Scored on ten worlds, every one of these archetypes scores below the reference in at least 96\% of the sets of worlds (Fig.~\ref{fig:toy_worlds}).

The dependence on the number of sealed worlds $N$ is shown in Figs.~\ref{fig:toy_nworlds} and~\ref{fig:toy_chance}. The median of the ratio $R$ of an example to the reference on the same worlds changes little with $N$, while its spread over the sets of worlds narrows. On one world, one of these examples outscores the reference in up to 35\% of the sets, on ten worlds in about 3\% or fewer, and on twenty in fewer than 0.5\%.

\begin{figure}[!t]\centering
\includegraphics[width=\colfigwidth]{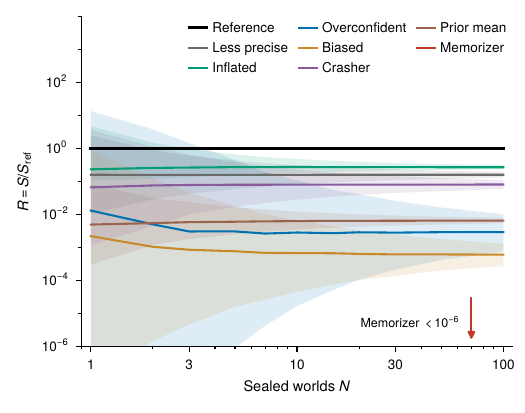}
\caption{Median and central 68\% interval of the ratio $R$ of each archetype in the toy of Fig.~\ref{fig:toy_reported} to the expert reference analysis on the same worlds, against the number of sealed worlds $N$.}
\label{fig:toy_nworlds}
\end{figure}

\begin{figure}[!t]\centering
\includegraphics[width=\colfigwidth]{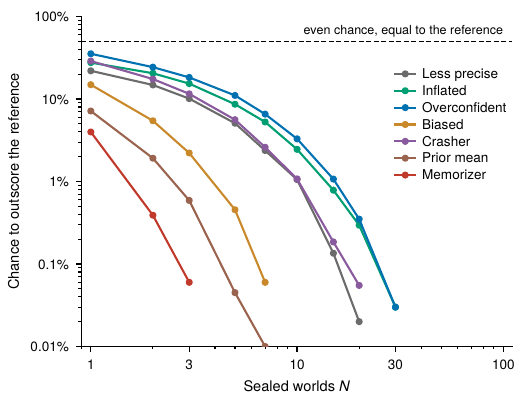}
\caption{Chance that an archetype outscores the expert reference analysis on the same worlds, over 20\,000 sets of $N$ sealed worlds, where an analysis as good as the reference would be at 50\%. A line ends where no set is left.}
\label{fig:toy_chance}
\end{figure}

\section{Task and harness settings}
\label{app:settings}

\subsection{Evaluation contract}
\label{app:contract}
The evaluation contract of Sec.~\ref{sec:evaluation} fixes the following.
\begin{itemize}[nosep, leftmargin=1.4em]
\item Submission: a directory whose pipeline produces a result on a world, in a fixed software environment and within limits of time and memory.
\item Parameters of interest: named signal strengths, cross sections, and masses and widths of the Higgs, $\mathrm{Z}$ and $\mathrm{W}$ bosons and of the top quark.
\item Nuisance parameters: named by source, the scale, resolution and efficiencies of the physics objects, pileup and theory.
\item Likelihood-fit tasks: the statistical model in the HistFactory format~\cite{histfactory,pyhf_joss} with the selected events, and for a mass also the likelihood on a grid of fixed masses.
\item Calibration tasks: the correction functions in the correctionlib format~\cite{correctionlib}.
\item Self-check: a script given to the agent runs the pipeline on the public world in the same way as the evaluation and checks the format of its outputs.
\end{itemize}

\subsection{Prompt tiers}
\label{app:prompts}
The prompts of all tasks share the same structure. The lowest tier, P0, just states the task and the output format. Each higher tier adds more guidance, such as the physics context, an ordered expert plan, or the sources of systematic uncertainty to evaluate, without specifying their size. No prompt tier mentions the expert reference analysis or states which detector effects vary between worlds. As an example, the three prompts of \hfl are given in full below.

\subsubsection{The \hfl prompts}
\label{app:prompts_hfl}
P0 is the first block, P1 adds the second and P2 the third. The evaluation contract of App.~\ref{app:contract}) is passed as a file reference \texttt{EVAL\_CONTRACT.md}.

\begin{nolinenumbers}
\begin{tcolorbox}[breakable, colback=black!4, colframe=black!4, boxrule=0pt, arc=2pt, left=6pt, right=6pt, top=5pt, bottom=5pt, fonttitle=\bfseries\small, coltitle=black, colbacktitle=black!10, title={P0}, before skip=6pt, after skip=6pt]
\promptfont
Perform a high-energy-physics analysis of the dataset described in DATASET.md.  The target of the analysis: the mass and the signal strength of the Higgs boson in the H$\to$ZZ$^*$$\to$4$\ell$ channel.

Deliver a reproducible analysis directory that conforms to EVAL\_CONTRACT.md: a \texttt{run.sh <dataset\_dir> <output\_dir>} entrypoint that writes \texttt{<output\_dir>/RESULT.json}.  If you report corrections such as lepton scale factors, give each as one constant -- its average over the population EVAL\_CONTRACT.md \S{}4 defines.
\end{tcolorbox}
\begin{tcolorbox}[breakable, colback=black!4, colframe=black!4, boxrule=0pt, arc=2pt, left=6pt, right=6pt, top=5pt, bottom=5pt, fonttitle=\bfseries\small, coltitle=black, colbacktitle=black!10, title={P1 adds}, before skip=6pt, after skip=6pt]
\promptfont
Physics envelope: a Higgs-like signal, if present, has $m_{\mathrm{H}}$ $\in$ [100, 200] GeV. Signal simulation is provided at $m_{\mathrm{H}}$ = 125 GeV only.

Physics context: the dataset is dominated by Drell--Yan; the backgrounds to a tight four-lepton selection are dominated by the irreducible ZZ$\to$4$\ell$ continuum, with reducible Z+X below it.  The data were recorded by a lepton-multiplicity trigger menu with prescaled lower-multiplicity paths (DATASET.md).  The detector response of the data may differ from the nominal simulation.  The published measurement in this channel under the same (2016) conditions is CMS, JHEP 11 (2017) 047, arXiv:1706.09936.

Tooling context: the field's standard stack applies -- columnar, vectorized analysis rather than per-event Python loops, and a proper statistical model for the inference.
\end{tcolorbox}
\begin{tcolorbox}[breakable, colback=black!4, colframe=black!4, boxrule=0pt, arc=2pt, left=6pt, right=6pt, top=5pt, bottom=5pt, fonttitle=\bfseries\small, coltitle=black, colbacktitle=black!10, title={P2 adds, and the sentence on the detector response ends with ``-- account for this''}, before skip=6pt, after skip=6pt]
\promptfont
Analysis plan -- follow it.  Take the published measurement's strategy where this dataset allows it; its lepton calibration methods are described in detail in CMS, Phys. Rev. D 89 (2014) 092007, arXiv:1312.5353.  Take methods from the literature, not numbers: every correction is measured on this dataset.

\begin{enumerate}[nosep, leftmargin=1.6em]
\item Selection: build 4$\ell$ candidates per event -- lepton identification, isolation and impact-parameter requirements, final-state-radiation photon recovery, then an opposite-sign same-flavour Z1 closest to m\_Z and a Z2 -- developed and validated on the simulation.
\item Corrections and systematic uncertainties: derive from the data, per lepton flavour and in bins of pT and |$\eta$|, every correction the published analysis applies to its simulation -- the momentum scale and resolution from the dilepton lineshapes, the efficiency of each step of the lepton selection with tag-and-probe -- and decide which of them differ from the nominal simulation here.  Apply them to the signal and background templates and yields, and carry their uncertainties, the luminosity uncertainty and the theory uncertainties of the predicted processes as nuisance parameters of the model.
\item Normalisation: expected yield of a simulated process = \texttt{effective\_xsec\_pb $\times$ lumi\_fb $\times$ 1000 $\times$ n\_selected/n\_preselection} (entries from each process's \texttt{meta.json}; the luminosity from \texttt{lumi.json}).  Data events are recorded through the trigger menu's prescales (DATASET.md).  The signal changes with $m_{\mathrm{H}}$ through its cross section $\times$ branching ratio (LHC Higgs Cross Section Working Group tables, relative to the provided 125 GeV point) and through its acceptance $\times$ efficiency, which the 125 GeV simulation alone does not give: derive it, e.g. from your own generator-level simulation (any generator may be installed in your environment).
\item Backgrounds: the irreducible ZZ from the simulation; the reducible Z+X from data control regions (lepton misidentification rates measured in data), with the simulation as a cross-check.
\item Fit: a binned likelihood of the four-lepton mass spectrum -- the signal template from the 125 GeV simulation, its peak tracking $m_{\mathrm{H}}$, with your corrections applied -- with $m_{\mathrm{H}}$ and $\mu$ floating, built in a statistical framework that can express a floating $m_{\mathrm{H}}$.
\item Uncertainties: profile-likelihood (or equivalent) intervals; report the statistical and systematic components and their quadrature total.
\item Validation: make sure the statistical model is appropriate -- the corrected simulation against the data in control regions (Z$\to$$\ell$$\ell$, Z+$\ell$, the four-lepton sidebands), toy-based goodness of fit (not just the asymptotic saturated test) and the post-fit nuisance impacts on $\mu$; iterate the model if any of them looks wrong.
\item Outputs: everything the contract specifies; self-check and fix until it passes.
\end{enumerate}

Track your work for reproducibility: develop the submission as a git repository with meaningful commits as the analysis evolves, and keep a log of the decisions taken (selection choices, corrections, model iterations) alongside the code -- the analysis must be auditable end to end.

Practical notes: read branches selectively with uproot and keep the event processing columnar/vectorized (awkward) -- a per-event Python loop over these datasets will not finish inside the run budgets of EVAL\_CONTRACT.md.
\end{tcolorbox}
\end{nolinenumbers}

\newcommand{\profilesfigure}{%
\begin{subfigure}{0.5\columnwidth}\centering
\includegraphics[width=\textwidth]{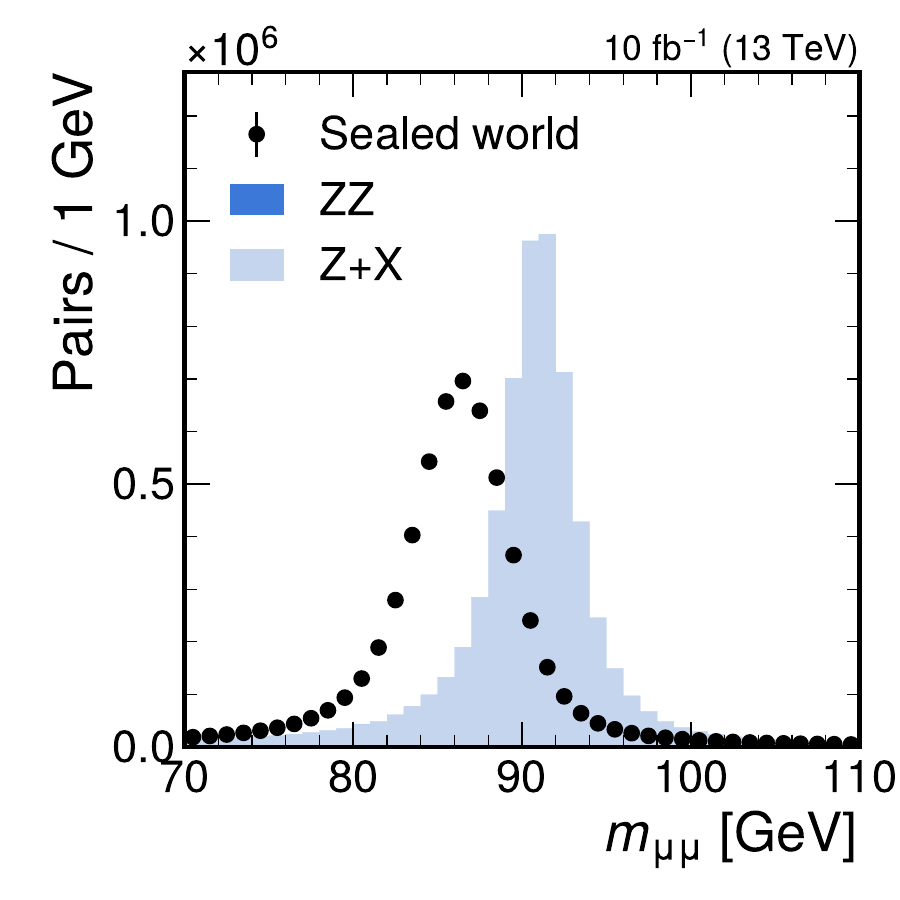}
\caption{}\label{fig:profiles_mmumu}
\end{subfigure}%
\begin{subfigure}{0.5\columnwidth}\centering
\includegraphics[width=\textwidth]{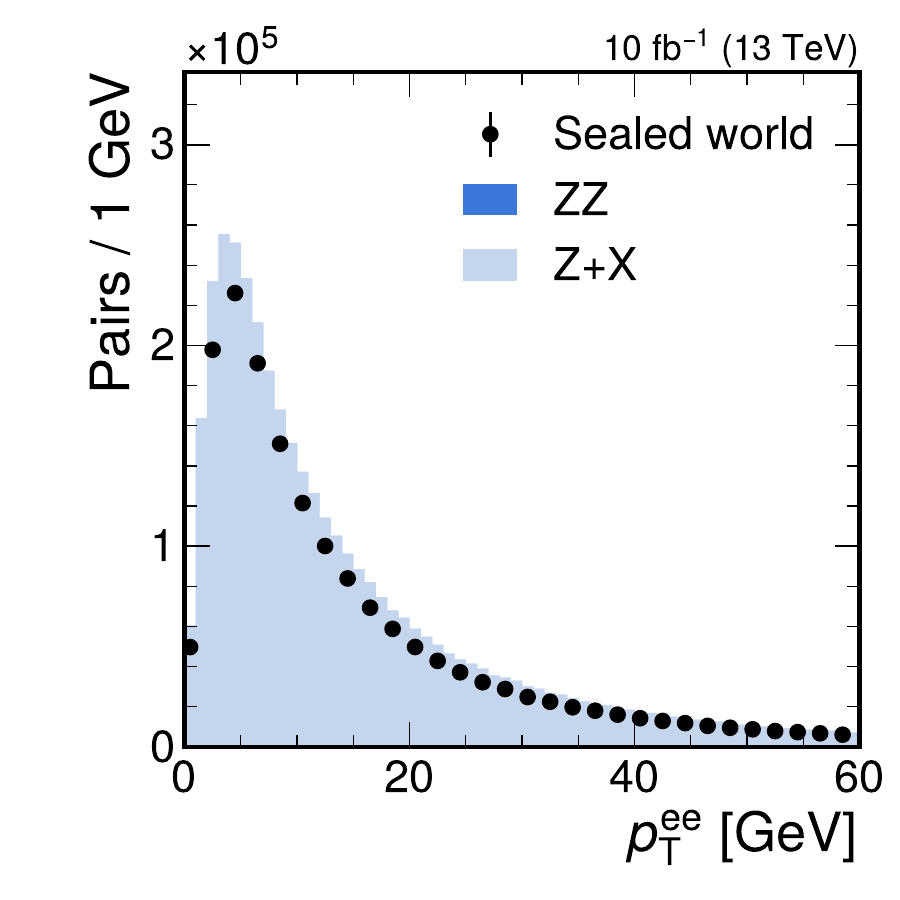}
\caption{}\label{fig:profiles_ptee}
\end{subfigure}

\begin{subfigure}{0.5\columnwidth}\centering
\includegraphics[width=\textwidth]{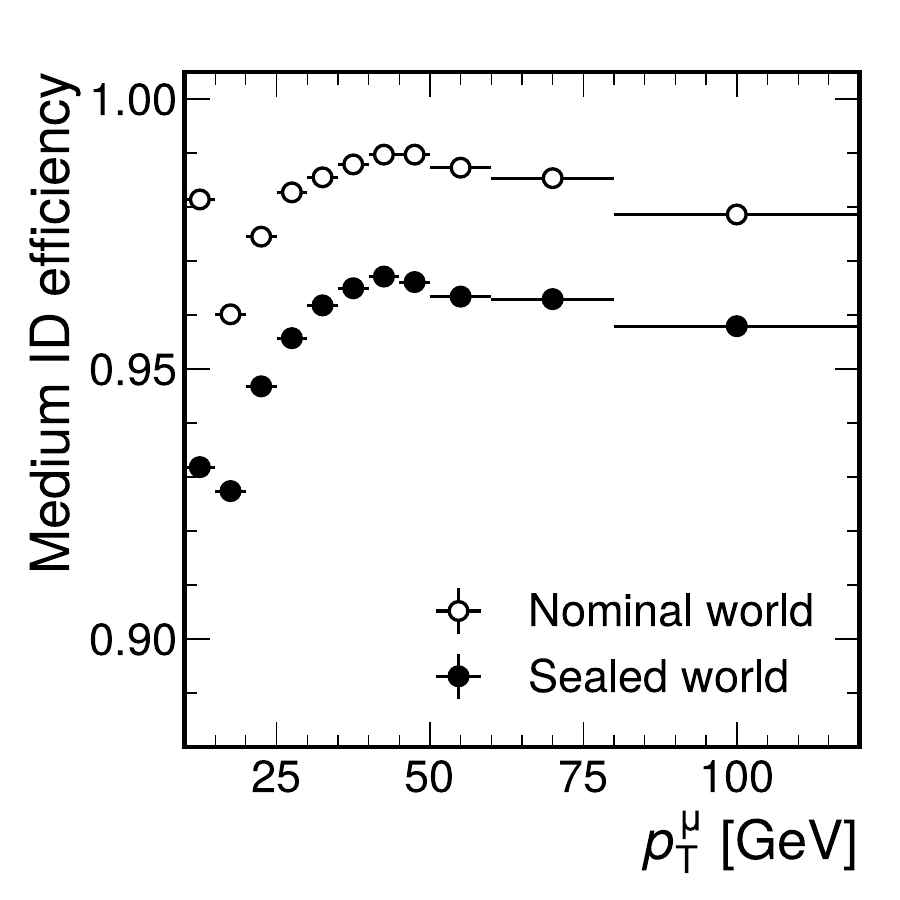}
\caption{}\label{fig:profiles_mu_id}
\end{subfigure}%
\begin{subfigure}{0.5\columnwidth}\centering
\includegraphics[width=\textwidth]{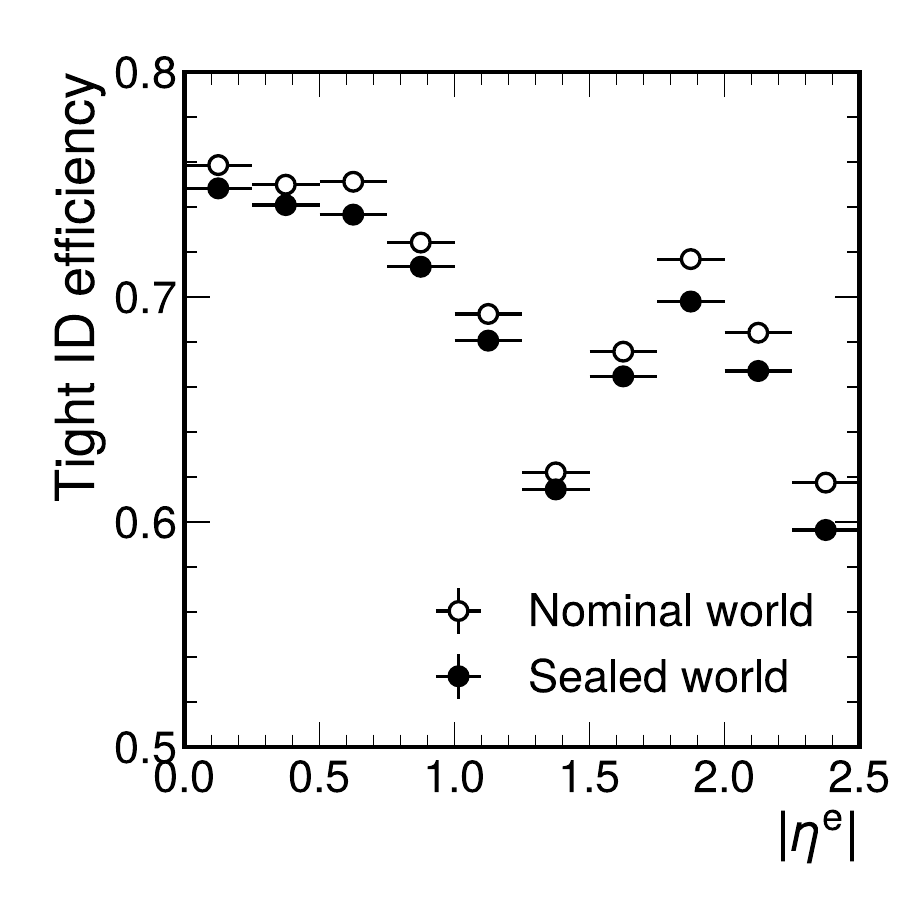}
\caption{}\label{fig:profiles_el_id}
\end{subfigure}
\caption{Injected detector effects in one example sealed \hfl world, with the largest modification of the momentum scale and resolution and lowered efficiencies. The dimuon mass at the $\mathrm{Z}$ peak (a) and the transverse momentum of the \Pe pair (b) in the sealed world (points) are shown over the nominal simulation as stacked histograms, with $\mathrm{Z}+\mathrm{X}$ the background with misidentified leptons. The tag-and-probe efficiencies~\cite{cms_tagprobe_2011} of the medium identification (ID) of \Pmu (c) and of the tight ID of \Pe (d) are shown for the nominal world, which is the public world on which analyses are developed (open markers), and for the sealed world (filled markers). The modified response shifts and widens the $\mathrm{Z}$ peak, and the lowered efficiencies reduce the number of selected pairs.}
\label{fig:profiles}}

\section{Datasets}
\label{app:datasets}

\subsection{Injected detector effects}
\begin{figure}[!t]\centering\profilesfigure\end{figure}
The twelve modified detector profiles combine four modifications of the momentum scale and resolution with three modifications of the efficiencies. A profile acts on every particle of the particle-flow reconstruction of CMS, which distinguishes \Pmu, \Pe, \Pgamma, charged hadrons and neutral hadrons~\cite{cms_pf_2017}. In order to maintain a self-consistent event, jets are reclustered, the missing transverse momentum is recomputed, and the links between jets, their constituents and secondary vertices are rebuilt after each modification. When the efficiencies are modified, each family of nested working points, such as the loose, medium and tight identification, changes consistently. The modified samples are then converted to NanoAOD~\cite{nanoaod}, the compact analysis format of CMS. Figure~\ref{fig:profiles} shows the two kinds of injected detector effects in an example sealed \hfl world. The \hwidth worlds take six of the modified profiles, applied to the leptons when a world is assembled.

In \jetcalib, the detector effects are injected in the particles of the jets. The response to each kind of particle is modified and smeared, in some worlds as a function of $\pt$ or $\eta$ or only in one region of the detector, and the jets are clustered again. The jet response then depends on the composition, the $\pt$ and the $\eta$ of the jet, as it does in data.

\subsection{Simulated samples}
The backgrounds of \hfl, \hwidth and \lepcalib are simulated Drell--Yan~\cite{cms_od_dy}, $\mathrm{q\bar{q}}\to\mathrm{ZZ}$~\cite{cms_od_zz}, $\mathrm{gg}\to\mathrm{ZZ}$~\cite{cms_od_ggzz_4e,cms_od_ggzz_4mu,cms_od_ggzz_2e2mu} and top quark pair~\cite{cms_od_tt} events. In \hfl and \hwidth, the Drell--Yan events are complemented with samples binned in the number of jets~\cite{cms_od_dy_jets} and in the scalar sum of the jet transverse momenta~\cite{cms_od_dy_ht}, and in \hwidth the $\mathrm{gg}\to\mathrm{ZZ}$ continuum is part of the off-shell samples described below. In the CMS Open Data samples, Drell--Yan events are generated with MadGraph5\_aMC@NLO~\cite{madgraph}, with jets merged in the MLM~\cite{Alwall:2007fs} or the FxFx~\cite{Frederix:2012ps} scheme, $\mathrm{q\bar{q}}\to\mathrm{ZZ}$ and top quark pair events with POWHEG~\cite{Nason:2004rx,Frixione:2007vw,Alioli:2010xd}, $\mathrm{gg}\to\mathrm{ZZ}$ events with MCFM~\cite{mcfm}, and multijet events with Pythia~8~\cite{pythia82}. Parton showering and hadronization are simulated with Pythia~8 with the CP5 tune~\cite{CMS:2019csb}, which the private samples use as well. The off-shell samples use the PDF4LHC15 parton distribution functions~\cite{Butterworth:2015oua}.

\subsection{Off-shell samples}
\label{app:width}
The off-shell samples of \hwidth are generated at leading order with MadGraph5\_aMC@NLO~3.6.7~\cite{madgraph}: gluon fusion, $\mathrm{gg}\to\mathrm{ZZ}\to4\ell$, loop induced and normalized with a QCD correction that depends on the four-lepton mass, and electroweak production, $\mathrm{qq}\to4\ell\mathrm{qq}$, at tree level. The Higgs-boson signal, the continuum and their sum with the interference are generated separately, and the width is changed through the couplings of the Higgs boson so that the on-shell rate stays fixed, as in the CMS off-shell measurement~\cite{cms_offshell_2022}. The off-shell events of a world are generated at its width, since a reweighting of events at the SM width that depends only on the four-lepton mass would not reproduce the distribution of a kinematic discriminant. In some sealed worlds, the QCD correction to $\mathrm{gg}\to\mathrm{ZZ}$ production is also varied within its scale uncertainty as a function of the four-lepton mass, so that the theory prediction differs from that of the nominal simulation. Below a four-lepton mass of 130\GeV, the worlds contain the continuum and the on-shell signal from Pythia~8~\cite{pythia82}. All samples are passed through the official CMS simulation of the 2016 conditions~\cite{cms_od_premix}, and the agent receives them at the SM width only.

\section{Tasks and expert reference analyses}
\label{app:closure}

Each task has an expert reference analysis that sets the precision and wall time benchmarks for the model evaluation. It runs under the same contract as a submission, on the same inputs and without access to the truth. Every reference recovers every injected truth without bias and with pulls of unit width, within the statistical resolving power of the $N$ worlds.

\subsection{\hfl}
\label{app:ref_hfl}

\textit{Task.} \hfl measures the mass $\mh$ of the Higgs boson and its signal strength $\mu$ in the decay $\hzz$ channel, with $\ell=\Pe,\Pmu$. The signal strength is defined as the production rate in units of the Standard Model (SM) prediction. The task requires lepton calibration, an estimate of the background from misidentified leptons, a mass-dependent signal model, a likelihood fit and systematic uncertainties. As in a real experiment, the lepton response of a world may differ from the nominal simulation and, left uncorrected, bias the extracted mass and signal strength. The leptons must therefore be calibrated with the $\Zll$ events of the same world and the resulting calibration must be propagated to the four-lepton signal region. With its very small background, the four-lepton channel provides the tightest single-channel constraint on its mass, $\mh=125.04\pm0.12\GeV$ with 138\ifb of CMS Run-2 data~\cite{cms_mass_width_4l_2024}.

\begin{figure}[!htb]\centering
\includegraphics[width=\colfigwidth]{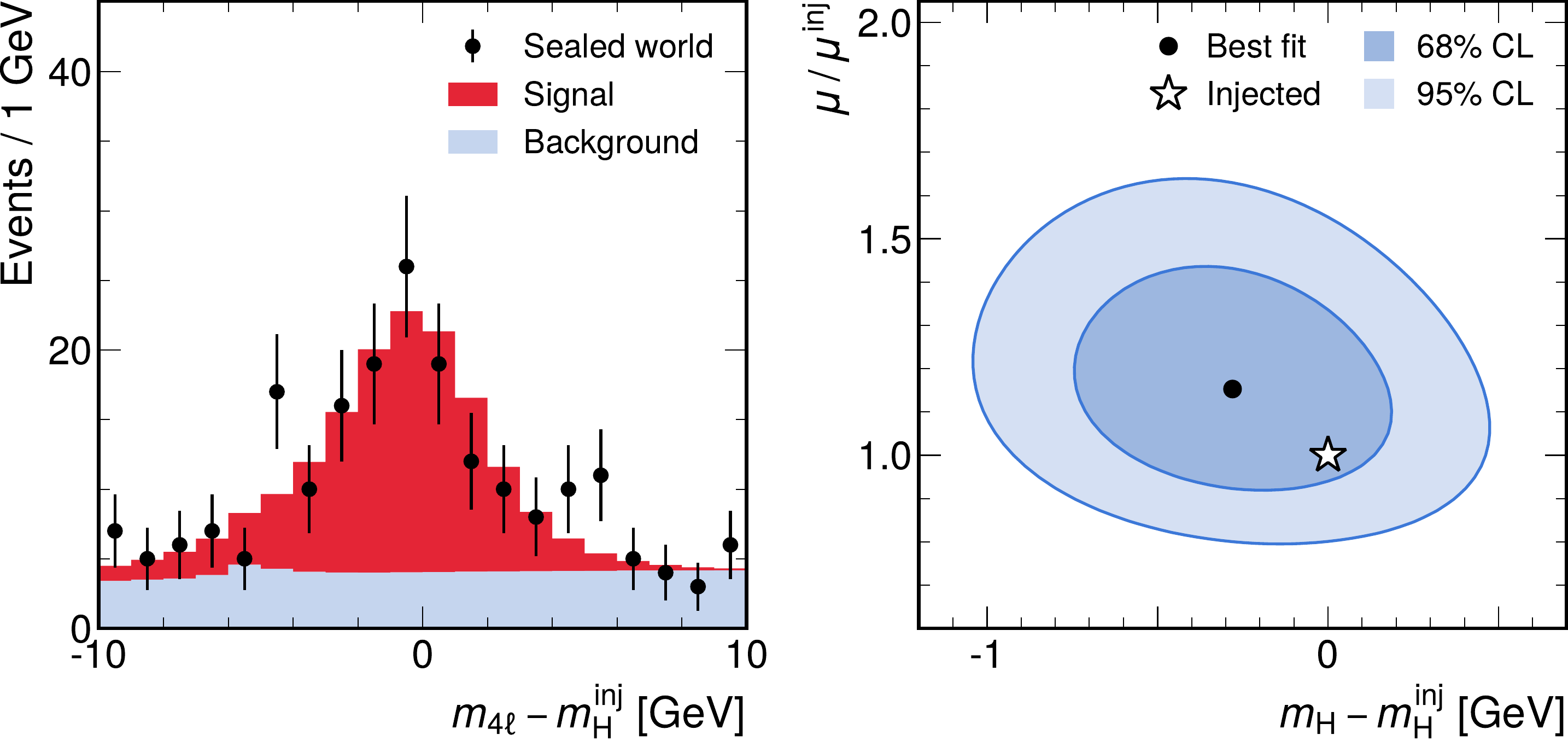}
\caption{Expert reference analysis of \hfl on one sealed world: the fitted four-lepton mass relative to the injected $\mh$ (left) and the 68\% and 95\% confidence regions in $\mh$ and $\mu$ relative to the injected values (right).}
\label{fig:obs_hfl}\label{fig:ref_hfl}
\end{figure}

The \hfl reference follows the published CMS measurement with 35.9\ifb of Run-2 data~\cite{cms_hzz4l_2016}.

\textit{Lepton calibration.} The momentum scale and resolution of each flavor are measured on $\Zll$ events. The simulated leptons are scaled and smeared in categories of $\pt$ and $|\eta|$ of the two leptons, and the dilepton mass templates are fitted to the pseudodata, with the procedure repeated until the corrections are stable. The angular resolutions are fitted in the same events, from the width of the mass peak and the acoplanarity of the pair. The mass uncertainty of each event is calibrated on the same events, in bins of the predicted uncertainty.

\textit{Efficiencies.} The efficiency of the full lepton selection is measured by tag-and-probe~\cite{cms_tagprobe_2011} in pseudodata and simulation, with a double-sided Crystal Ball function for the signal. The systematic uncertainty is the shift under an alternative signal and background model.

\textit{Selection.} Four-lepton candidates are selected following the criteria of the CMS analysis, requiring recovered final-state radiation, $40<m_{\mathrm{Z}_1}<120\GeV$ and $12<m_{\mathrm{Z}_2}<120\GeV$, and a kinematic refit of the $\mathrm{Z}_1$ mass. The events are split by final state, $4\Pmu$, $4\Pe$ and $2\Pe2\Pmu$, and by the relative mass uncertainty.

\textit{Backgrounds.} The background from misidentified leptons is estimated from control regions of the pseudodata, using opposite-sign and same-sign methods and then combining the distributions. The misidentification rates are measured in $\mathrm{Z}+\ell$ events for light flavor and in $\Pe\Pmu+\ell$ events for heavy flavor, and the prompt contribution is subtracted. The $\mathrm{ZZ}$ backgrounds are taken from simulation, and the normalization of $\mathrm{q\bar{q}}\to\mathrm{ZZ}$ is left free and constrained by a control region at $140<\mfl<165\GeV$.

\textit{Signal model.} The signal in each channel is a double-sided Crystal Ball function whose shape scales with $\mh$. The cross sections and branching fractions are those of Ref.~\cite{lhchxswg_yr4}, and the dependence of the acceptance on $\mh$ comes from a simulation at generator level~\cite{pythia82}.

\textit{Fit.} The likelihood is constructed using $\mfl$ bins ranging from 105 to 140\GeV with a 0.5\GeV width, five bins in the mass uncertainty and three final states, and is maximized using~\cite{pyhf_joss}. The $\mh$, $\mu$ and the $\mathrm{q\bar{q}}\to\mathrm{ZZ}$ normalization are treated as free parameters in the fit. The nuisance parameters are the lepton scale and resolution, the efficiencies, the misidentified-lepton background, the luminosity and the theory normalizations. The profile likelihood sets the intervals, and their statistical part is obtained with the constrained nuisance parameters fixed. The goodness of fit is taken from pseudoexperiments with a saturated-likelihood statistic~\cite{baker_cousins_1984}.

\textit{Closure.} The reference passes closure on the 40 sealed worlds of \hfl. The pull distribution of $\mh$ includes only worlds containing a Higgs signal. Figure~\ref{fig:ref_hfl} shows the fitted spectrum on one of them and the delivered likelihood.

\subsection{\hwidth}
\label{app:ref_hwidth}

\textit{Task.} \hwidth measures the total width $\GH$ of the Higgs boson as the ratio $\xi=\GH/\GSM$ to the SM prediction $\GSM\approx4.1\MeV$~\cite{lhchxswg_yr4}, and its on-shell signal strength $\mu_{\mathrm{on}}$ in the decay $\mathrm{H}\to\mathrm{ZZ}\to4\ell$ with $\ell=\Pe,\Pmu$. Since the width lies far below the detector resolution, it is measured from off-shell production above the $\mathrm{ZZ}$ threshold~\cite{caola_melnikov_2013}. With the on-shell rate fixed, the off-shell signal grows with $\mu_{\mathrm{on}}\xi$ and its interference with the $\mathrm{gg}\to\mathrm{ZZ}$ and electroweak $\mathrm{ZZ}$ continuum with $\sqrt{\mu_{\mathrm{on}}\xi}$. The task requires an analysis with lepton calibration, a joint fit of the on-shell peak and the off-shell spectrum, a kinematic discriminant against the continuum, from which the signal differs only in its kinematics, and a prediction at any width built from simulation at the SM width alone. As in \hfl, the leptons must be calibrated with the $\Zll$ events of the same world. A width above the SM prediction would reveal decays beyond the SM. The measurement from off-shell production gives $\GH=2.9^{+2.3}_{-1.7}\MeV$ with 138\ifb of CMS Run-2 data in the four-lepton final state~\cite{cms_mass_width_4l_2024}.

\begin{figure}[!htb]\centering
\includegraphics[width=\colfigwidth]{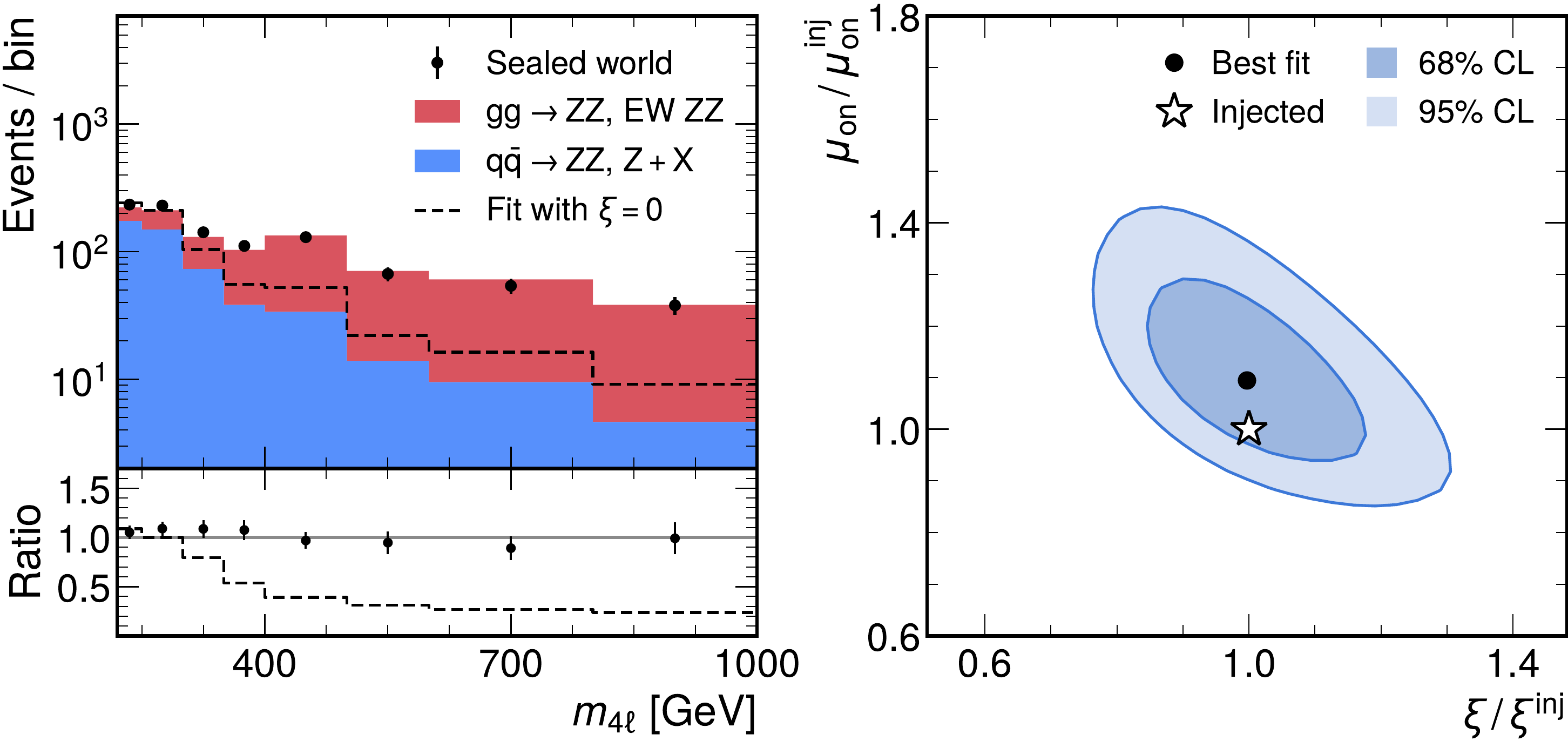}
\caption{Expert reference analysis of \hwidth on one sealed world: the fitted four-lepton mass above 220\GeV at the highest values of the kinematic discriminant (left) and the 68\% and 95\% confidence regions in $\xi$ and $\mu_{\mathrm{on}}$ relative to the injected values (right).}
\label{fig:obs_hwidth}\label{fig:ref_hwidth}
\end{figure}

The \hwidth reference follows the CMS off-shell measurement~\cite{cms_offshell_2022} and takes the lepton calibration, the selection, the background from misidentified leptons and the on-shell signal model of the \hfl reference.

\textit{Regions.} The on-shell events at $105<\mfl<140\GeV$ are split further into five bins of a kinematic discriminant against the $\mathrm{ZZ}$ continuum, and above 220\GeV, $\mfl$ is fitted in three bins of a second discriminant for each final state, with a control region at $165<\mfl<220\GeV$. Both discriminants are learned from the simulation in place of the matrix-element discriminants of CMS.

\textit{Model.} The prediction at any width is $T(g)=B+\sqrt{g}\,(\mathit{SBI}-S-B)+g\,S$ with $g=\mu_{\mathrm{on}}\xi$ and $S$, $B$, $\mathit{SBI}$ the signal, the continuum and their sum with the interference at the SM width, for gluon fusion and electroweak production separately. One binned likelihood of both regions is maximized with $\xi$, $\mu_{\mathrm{on}}$ and the $\mathrm{q\bar{q}}\to\mathrm{ZZ}$ normalization free, with nuisance parameters for the luminosity, the lepton efficiencies, the misidentified-lepton background, the theory of gluon fusion and the statistical uncertainty of the simulation. The 68\% confidence set of $\xi$ is the union of the intervals with $q(\xi)=-2\ln\frac{L(\xi)}{L_{\max}}\le1$ in a scan from 0 to 100.

\textit{Closure.} The reference passes closure on the 30 sealed worlds of \hwidth. Figure~\ref{fig:ref_hwidth} shows the fitted off-shell spectrum on one of them and the delivered likelihood.

\subsection{\lepcalib}
\label{app:ref_lepcalib}

\textit{Task.} \lepcalib calibrates the momentum scale, the momentum and angular resolutions, and the identification and trigger efficiencies of \Pe and \Pmu. The differences between data and simulation are described by corrections, each a function of the lepton kinematics, which analyses apply to simulated events~\cite{bodek_muon_scale_2012,cms_muon_2018,cms_egamma_2021}. In \lepcalib these corrections themselves are the result. The precision of the lepton calibrations limits that of measurements, and the uncertainty of the lepton momentum scale dominates the systematic uncertainty of the CMS measurement of the Higgs-boson mass in Run-2~\cite{cms_hzz4l_2016}. Leptons are calibrated with $\mathrm{Z}$-boson decays. The dilepton mass distribution fixes the momentum scale and resolution, and the tag-and-probe method measures efficiencies~\cite{cms_tagprobe_2011}.

\begin{figure}[!htb]\centering
\includegraphics[width=\colfigwidth]{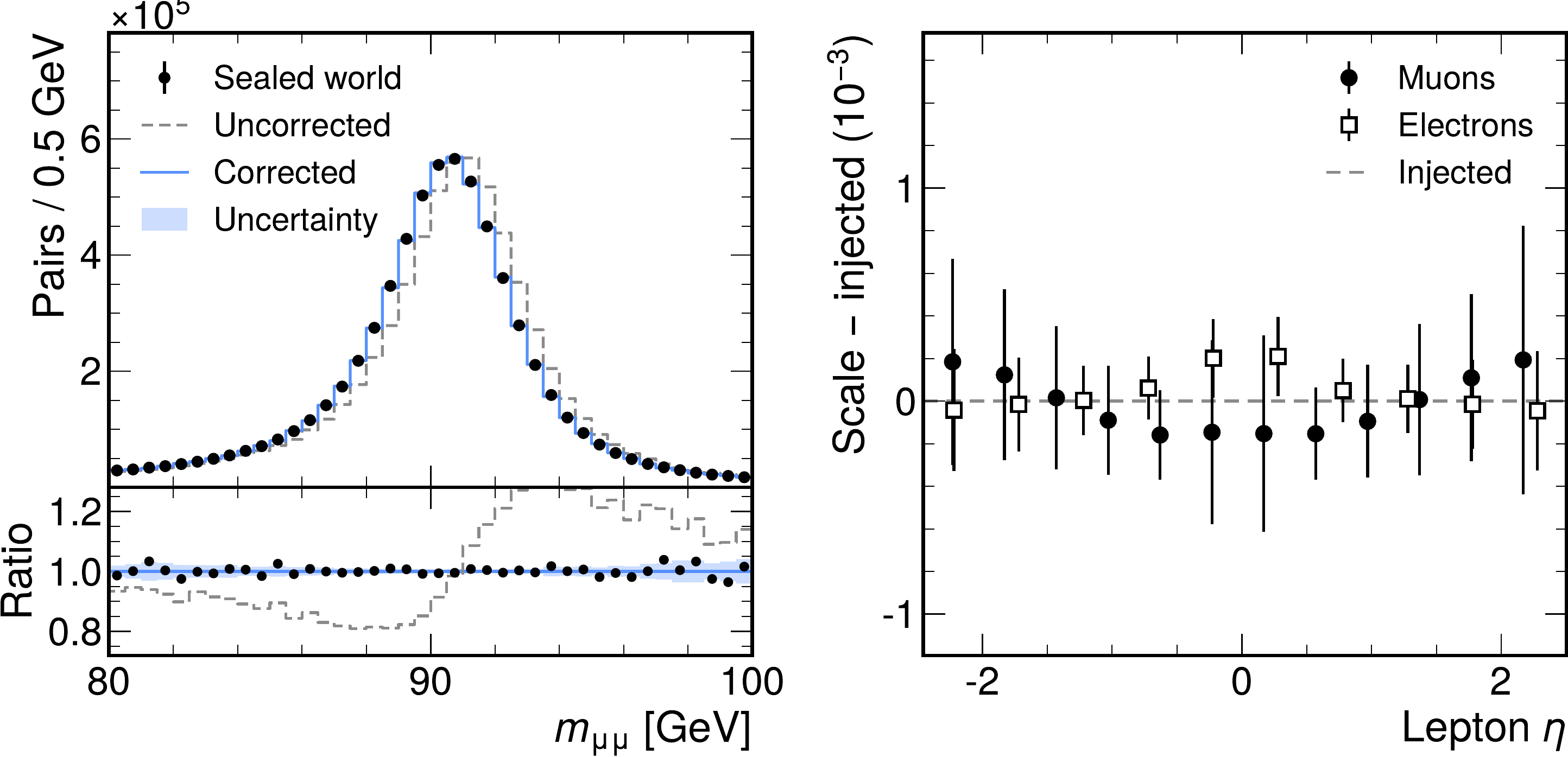}
\caption{Expert reference analysis of \lepcalib on one sealed world: the dimuon mass at the $\mathrm{Z}$ peak before and after the corrections (left) and the momentum-scale correction relative to the injected one (right).}
\label{fig:obs_lepcalib}\label{fig:ref_lepcalib}
\end{figure}

The \lepcalib reference follows the CMS lepton calibrations~\cite{bodek_muon_scale_2012,cms_muon_2018,cms_egamma_2021,cms_tagprobe_2011}. It calibrates each flavor by forward folding and measures the efficiencies with tag-and-probe.

\textit{Scale and resolution.} One opposite-sign pair of identified, isolated leptons per event is used. The simulated leptons are scaled by a polynomial in $\ln\pt$ and $|\eta|$ and smeared in $\pt$, $\eta$ and $\phi$ by an extra variance per lepton, and the selection is applied after the folding. Simulation and pseudodata are compared in categories of the two leptons. The truncated mean of $\ln\mll$ determines the scale and its truncated variance the widths. The $\phi$ resolution is determined by the shape of the acoplanarity $\pi-|\Delta\phi|$ at fixed lepton $\pt$. The $\eta$ resolution is determined for muons by the shape of the $|\eta|$ distribution at the edge of the acceptance, and for electrons by the difference between the $\eta$ of the calorimeter cluster and of the track. A single least-squares fit gives scale and widths, with the statistical uncertainty of the simulated templates taken into account.

\textit{Efficiencies.} Tag-and-probe pairs have a tag that is identified, isolated and matched to the single-lepton trigger. The efficiencies are measured after the momentum calibration, in bins of $\pt$ and $|\eta|$, for each requested identification and trigger leg. The ratio of pseudodata to simulation is described by a polynomial in $1/\pt$ and $|\eta|$, fitted to the bins by a binomial maximum likelihood.

\textit{Functional forms.} The forms are chosen from the pseudodata. A constant is fitted first, and the next polynomial degree is then kept while a likelihood-ratio test prefers it at the 5\% level, for the scale, the widths and each efficiency.

\textit{Uncertainties.} Each correction has several associated uncertainty sources, which are treated as fully correlated across the kinematic range. The statistical uncertainties of pseudodata and simulation are propagated from the fits. The systematic sources are a narrower $\mathrm{Z}$ window and a higher tag threshold for the efficiencies, narrower fit windows for the calibration, and for every form the shift to the adjacent degree.

\textit{Closure.} On the ten sealed worlds of the campaign, the reference agrees with the injected truths within its uncertainties. Figure~\ref{fig:ref_lepcalib} shows the $\mathrm{Z}$ peak after the corrections and the momentum scale on one of them.

\subsection{\jetcalib}
\label{app:ref_jetcalib}

\textit{Task.} \jetcalib calibrates the energy scale and the energy resolution of jets, hadronic objects that are difficult to handle at the LHC but enter most of its analyses. As in \lepcalib, the corrections and their uncertainties make up the result. The worlds hold $\mathrm{Z}(\to\Pmu\Pmu)+\mathrm{jet}$ and multijet events, the latter recorded with prescaled jet triggers. Unlike in \lepcalib, there is no resonance to calibrate on. The uncertainty of the jet energy scale limits the precision of many measurements. In CMS, the jet energy is corrected in steps, and the residual corrections for the differences between data and simulation are measured in data with the transverse-momentum balance of dijet, $\mathrm{Z}+\mathrm{jet}$ and $\Pgamma+\mathrm{jet}$ events~\cite{cms_jec_8tev_2017}. The pipeline reports the jet energy scale and the jet $\pt$ resolution of the pseudodata relative to the simulation, as functions of the jet $\pt$, $\eta$ and $\phi$ for $\pt>30\GeV$ and $|\eta|<2.5$. The balance is biased by additional radiation, by the resolution and by the trigger thresholds, and the forward regions are reached only through the balance of dijet events.

\begin{figure}[!htb]\centering
\includegraphics[width=\colfigwidth]{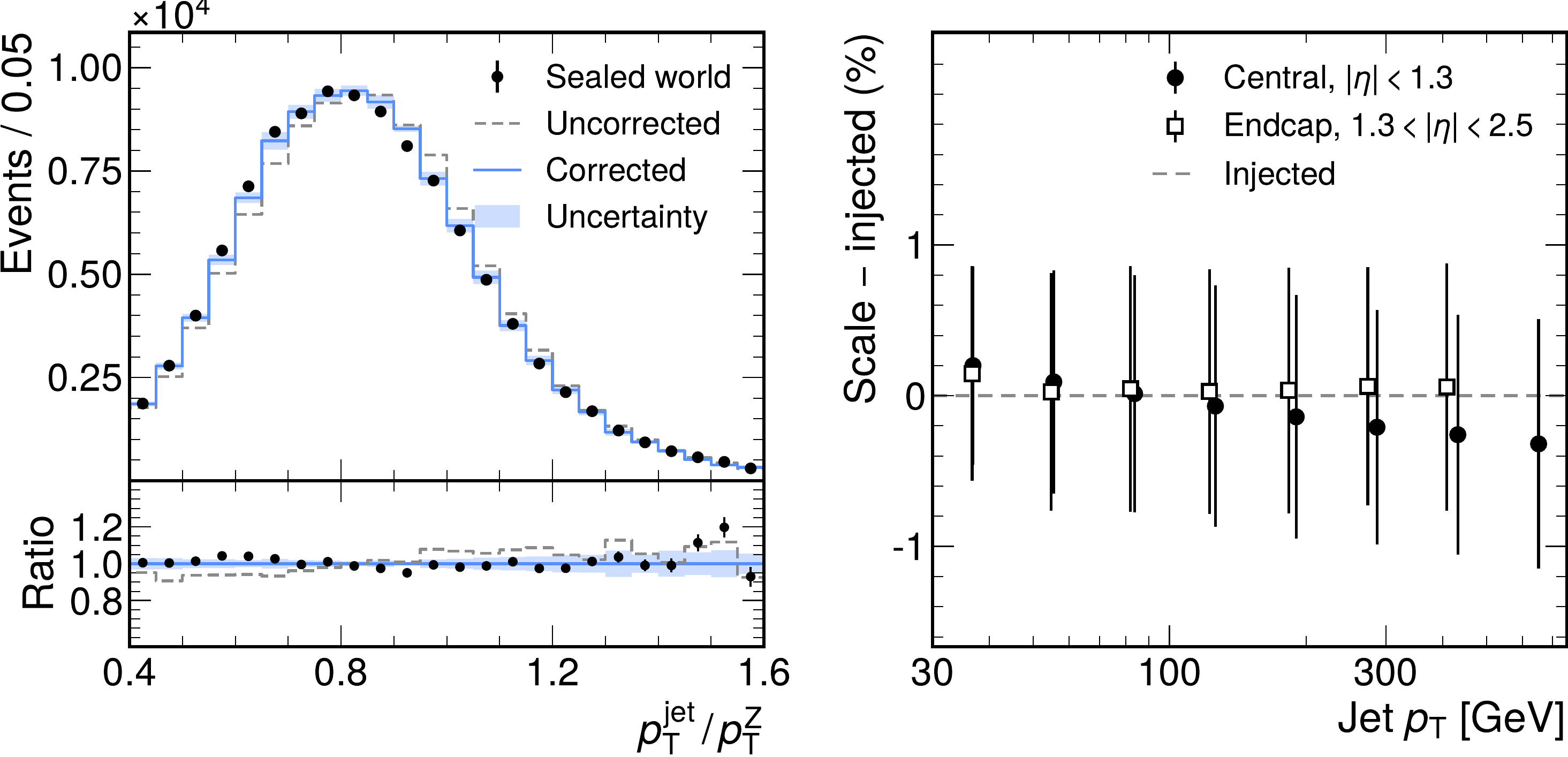}
\caption{Expert reference analysis of \jetcalib on one sealed world: the $\mathrm{Z}+\mathrm{jet}$ balance before and after the corrections (left) and the jet energy scale correction relative to the injected one (right).}
\label{fig:obs_jetcalib}\label{fig:ref_jetcalib}
\end{figure}

The \jetcalib reference follows the CMS calibration~\cite{cms_jec_8tev_2017} in factorized steps.

\textit{Simulated response.} The jets are corrected to the particle level with the response of the simulation, the mean ratio of reconstructed to particle-level $\pt$ in slices of $|\eta|$, fitted as a function of $\pt$ and inverted. The correction is applied to pseudodata and simulation alike. No pileup offset is needed, since pseudodata and simulation share the same pileup. Every change of a jet is propagated to the missing transverse momentum.

\textit{Resolution.} The extra resolution is measured from the width of the dijet asymmetry in back-to-back events, in bins of $|\eta|$ and of the average $\pt$, with a central reference jet for the forward bins. The difference of the squared widths in pseudodata and simulation is extrapolated to no additional jet activity, and fitted in four regions of $|\eta|$, the barrel, two endcap regions and the forward calorimeter.

\textit{Relative scale.} The dependence of the scale on $\eta$ is measured from the balance of dijet events with the missing-transverse-momentum projection fraction (MPF) method, with a central tag jet, as the ratio of pseudodata to simulation extrapolated to no additional jet activity. It is described in each $|\eta|$ bin by a constant and a term in $\ln\pt$.

\textit{Absolute scale.} The absolute scale in the central region is obtained from one global fit of the MPF response in $\mathrm{Z}(\to\Pmu\Pmu)+\mathrm{jet}$ events, the balance of multijet events, which extends the range in $\pt$, and the energy fractions of charged hadrons, neutral hadrons and photons in the jets. The model gives each particle species its own energy scale, linear in $\ln\pt$, so that the composition of the jets constrains the dependence of the response on $\pt$.

\textit{Uncertainties.} The statistical uncertainties of pseudodata and simulation are from a Poisson bootstrap of the events, with every measurement repeated on each replica. The systematic sources are the choices of the method, namely the extrapolation and the truncation of the asymmetry, the forms of the fits, the propagation of the resolution into the scale, and the dependence of the response on the jet flavor, measured in $\mathrm{Z}+\mathrm{jet}$ events.

\textit{Closure.} On the 32 sealed worlds, the reference agrees with the injected truths within its uncertainties. Figure~\ref{fig:ref_jetcalib} shows the balance after the corrections on one of them and the scale.

\end{document}